\documentclass[aps,prx,reprint,superscriptaddress,twocolumn,longbibliography]{revtex4-2}
\usepackage[utf8]{inputenc}
\usepackage{graphicx}
\usepackage{amsmath}
\usepackage{amsfonts}
\usepackage{amssymb}
\usepackage{bbm}
\usepackage[usenames,dvipsnames,svgnames,table]{xcolor}
\usepackage[USenglish]{babel}
\usepackage{hyperref}
\usepackage{grffile}
\usepackage{xcolor}
\usepackage{bm}
\usepackage{tabularx}
\usepackage{array}
\usepackage{mathtools}
\usepackage{makecell}
\usepackage{booktabs}
\usepackage{blindtext}

\usepackage{environ}
\usepackage{siunitx}
\usepackage{tikz}
\usetikzlibrary{positioning}
\usepackage{ifthen}
\usepackage{tikz-3dplot}
\usetikzlibrary{fadings}

\newcommand{\ignore}[1]{}

\newcolumntype{P}[1]{>{\centering\arraybackslash}p{#1}}

\allowdisplaybreaks

\newcommand{\ket}[1]{\left| #1 \right\rangle}

\newcommand{\noline}{$0$}
\newcommand{\bbone}{\mathbb{I}}

\begin{document}
\title{ Classification of Weyl points and nodal lines\\
based on magnetic point groups for spin-$\frac{1}{2}$ quasiparticles}
\author{Andy Knoll}
\email{andy.knoll@tu-dresden.de}
\author{Carsten Timm}
\email{carsten.timm@tu-dresden.de}
\affiliation{Institute of Theoretical Physics, Technische Universit\"at Dresden, 01062 Dresden, Germany}
\affiliation{W\"urzburg--Dresden Cluster of Excellence ct.qmat, Technische Universit\"at Dresden, 01062 Dresden, Germany}

\begin{abstract}
Symmetry-protected topological semimetals are at the focus of solid-state research due to their unconventional properties, for example, regarding transport. By investigating local two-band Bloch Hamiltonians in the spin-1/2 basis for the 122 magnetic point groups, we classify twofold-degenerate band touchings such as Weyl points, robust nodal lines on axes and in mirror planes, and fragile nodal lines. We find that all magnetic point groups that lack the product of inversion and time-reversal symmetries can give rise to topologically nontrivial band touchings. Hence, such nodes are the rule rather than the exception and, moreover, do not require any complicated multiband physics. Our classification is applicable to every momentum in the Brillouin zone by considering the corresponding little group and provides a powerful tool to identify magnetic and nonmagnetic topological semimetals.
\end{abstract}

\maketitle

\section{Introduction}

Materials with topologically protected band-touching points and lines close to the Fermi energy display a vast variety of fascinating physical properties \cite{Yan_Felser_2017,Armitage_Mele_2018,Lv_Qian_2021}. Prominent representatives are Weyl and nodal-line semimetals. Weyl semimetals are characterized by zero-dimensional band-touching points called Weyl points that are twofold degenerate. Weyl points are monopoles of the Berry curvature with positive or negative charge (chirality) \cite{Fang_Nagaosa_2003,Wan_Turner_2011}. Since there is no net flux of Berry curvature into or out of the first Brillouin zone the total chirality of all Weyl points always vanishes \cite{Armitage_Mele_2018,Nielsen_Ninomiya_1983}. Each Weyl point is then topologically protected by a nonzero Chern number \cite{Armitage_Mele_2018}, where the annihilation of Weyl points is only possible if two Weyl points of opposite chirality meet at the same position in momentum space \cite{Xu_Weng_2011,Burkov_Balents_2011,Neumann_Wigner_1929}. Weyl points can only be realized if at least one of time-reversal and inversion symmetry is absent because the product of the two symmetries implies that for any Weyl point there is another one with opposite chirality at the same momentum, i.e., there is in fact a Dirac point.

Early works on Weyl semimetals \cite{Halasz_Balents_2012,Burkov_Balents_2011,Zyuzin_Burkov_2012,Das_2013,Murakami_Iso_2007} employed models that sometimes seemed contrived and usually involved a four-dimensional Hilbert space describing the local degrees of freedom. This gave the impression that topologically protected band touchings are somehow exotic and only exist if complicated conditions are satisfied. However, ab initio calculations \cite{Xu_Weng_2011,Liu_Vanderbilt_2014,Hirayama_Okugawa_2015,Huang_Xu_Belopolski_2015,Weng_Fang_Bernevig_2015,Singh_Sharma_2012,Lu_Fu_2013,Shekhar_Nayak_2015} quickly made clear that Weyl points are a rather common feature of band structures. This is supported by our work, which shows that 19 out of the 32 monochromatic, 11 out of the 32 gray, and 23 out of the 58 dichromatic point groups can host Weyl points.

Compared to Weyl points, one-dimensional band-touch\-ing lines (nodal lines) rely on more stringent conditions. In the absence of spin-orbit coupling, fourfold degenerate Dirac nodal lines can be realized in systems with both time-reversal and inversion symmetry \cite{Fang_Weng_2016}, as discussed for graphene networks \cite{Weng_Liang_2015}, the antiperovskite Cu$_3$PdN \cite{Yu_Weng_2015}, Cu$_3$N \cite{Kim_Wieder_2015}, and Ca$_3$P$_2$ \cite{Xie_Schoop_2015}. However, the neglect of spin-orbit coupling is, at best, a valid approximation for light elements. Upon taking into account spin-orbit coupling, $\textrm{SU}(2)$ spin-rotation symmetry is broken, which causes bands with opposite spin to hybridize, so that a Dirac line node is either split into discrete points or gapped out \cite{Fang_Weng_2016}. In order to stabilize the Dirac line node in the presence of spin-orbit coupling, additional constraints such as nonsymmorphic \cite{Fang_Chen_2015, Meng_Liu_2020,yu2021encyclopedia} or off-centered symmetries \cite{Yang_Bojesen_2017} are necessary. For systems with broken time-reversal or inversion symmetry, twofold degenerate Weyl nodal lines can be stabilized by mirror symmetry \cite{Du_Bo_Wan_2017,Weng_Fang_Bernevig_2015,Chiu_Schnyder_2014}, in which case the hybridization is hindered by distinct mirror eigenvalues of the relevant bands.

Due to the variety of gapless topological phases of matter, it is apparent that systematic approaches to classify them are needed. The tenfold way, which classifies topological semimetals in terms of nonspatial symmetries (time-reversal symmetry, particle-hole symmetry, and chiral symmetry) \cite{Matsuura_Chang_2013, Zhao_Wang_2013, Chiu_Teo_2016}, and its extension to reflection-symmetry-protected semimetals \cite{Chiu_Schnyder_2014} reveal by which topological invariant a band touching is protected. However, these classifications make no statement about the existence or the location in momentum space of the band touchings, which can only be deduced from additional symmetries.

In addition to catalogs of emergent particles for the 230 space groups \cite{yu2021encyclopedia} and of topological materials \cite{Verginory_Elcoro_2019, Zhang_Jiang_2019}, various other studies provide symmetry-based classifications of different band touchings including Dirac and Weyl points \cite{Gao_Hua_2016,Yang_Nagaosa_2014, Fang_Gilbert_2012}, Kramers-Weyl points in chiral materials \cite{Chang_Wieder_2018}, nodal lines \cite{Gao_Hua_2016,Yamakage_Yamakawa_2016,Kobayashi_Yamakawa_2017,Xie_Gao_21}, and  three-, six-, and eightfold degenerate band-touching points giving rise to unconventional quasiparticles \cite{Bradlyn_Cano_2016}.
In all of these investigations, the presence of time-reversal symmetry plays a crucial role. Only recently, there have been more systematic studies to classify magnetic systems. For monochromatic point groups, Weyl-semimetallic phases of spinless particles have been identified using band-structure combinatorics \cite{Kruthoff_de_Boer_2017}. In Ref.~\cite{Yang_Fang_2021}, Yang \textit{et al.}~study band touchings in magnetic space groups that contain the product of time-reversal and inversion. Furthermore, with the aid of magnetic topological quantum chemistry \cite{Elcoro_Wieder_2021}, first-principles calculations have been performed to investigate the band touchings of magnetic compounds from the Bilbao Crystallographic Server \cite{Xu_Elcoro_2020}. For tetragonal magnetic space groups that are compatible with antiferromagnetic, ferrimagnetic, and ferromagnetic ordering, gapless phases and their boundary effects have been investigated \cite{Bouhon_Lange_2021,Lange_Bouhon_2021_boundary_effects}.

In this paper, we add to the systematization of band touchings by providing a classification of twofold-degenerate band-touching points and lines for the 122 magnetic point groups  \cite{Wigner_1931,Dimmock_Wheeler_1962,Cracknell_1965,Cracknell_Wong_1967,Bradley_Cracknell_1972,Bradley_Davies_1968,Bradley_Cracknell_1972,Erb_Hlinka_2020,bilbao_server}. We here consider each magnetic point group as a subgroup of $\textrm{SU}(2)\otimes\{e,i\}\otimes\{e,\Theta\}$, where $\Theta$ denotes time reversal and $i$ inversion. Hence, the internal degree of freedom is described by a two-dimensional Hilbert space with the basis $\{\ket{\uparrow},\ket{\downarrow}\}$, where $\ket{\uparrow}$ and $\ket{\downarrow}$ transform like a spin of length 1/2 under the magnetic point group. Then, for each magnetic point group a pseudospin-1/2 Bloch Hamiltonian around a reference point (RP) is constructed and used to investigate the twofold degenerate band touchings in the vicinity of that point. Our results are valid for all pairs of bands that are described by a pseudospin with the same transformation behavior as a true spin-1/2. Our analysis applies to the vicinity of any point in the Brillouin zone. In case of the $\Gamma$ point, the relevant symmetry group is the full magnetic point group of the structure. For other points, one has to choose the corresponding little group.

\section{Method}
\label{sec:method}

We begin by explaining the method used to classify twofold degenerate band touchings. In the vicinity of a RP $\bm{k}_0$, every pseudospin-1/2 Hamiltonian can be written as
\begin{equation}\label{eq:general_twoband_Hamiltonian}
	H(\bm{k})=f_0(\bm{k})\,\sigma_0+\sum_{i=x,y,z} f_i(\bm{k})\,\sigma_i ,
\end{equation}
where $\bm{k}$ is the momentum relative to the RP, $f_\mu(\bm{k})$ with $\mu=0,x,y,z$ are real-valued functions, $\sigma_0$ is the $2\times 2$ identity matrix, and $\sigma_i$ with $i=x,y,z$ are the Pauli matrices. Our strategy is to use representation theory to find the general form of the Hamiltonian $H(\bm{k})$ allowed by symmetry. $H(\bm{k})$ has to be an irreducible tensor operator belonging to the trivial irreducible representation (irrep) for monochromatic point groups, or to the trivial irreducible corepresentation (corep) for gray and dichromatic point groups. The matrices $\sigma_\mu$ can be classified as irreducible tensor operators of irreps or irreducible coreps. $\sigma_0$ of course belongs to the trivial irrep or corep. The transformation properties of $\sigma_i$ with $i=x,y,z$ can be found by noting that $\sigma_i$ are the components of the spin operator, up to a factor of $\hbar/2$. For monochromatic point groups, the transformation properties and thus the irreps are determined by viewing the corresponding structural point group as a subgroup of $\mathrm{SU}(2)$ \footnote{One then also finds that the two spin states {$|{\uparrow}\rangle$} and {$|{\downarrow}\rangle$} belong to one or two double-valued irreps. For the latter case, the spin states belong to different complex-conjugated double-valued irreps. Otherwise, they are either basis functions of a two-dimensional double-valued irrep, for example the irrep $E_{1/2}$ of the point group $T_d$, or they both belong to the same one-dimensional double-valued irrep.}. For gray and dichromatic point groups, the irreducible coreps of $\sigma_i$ are determined by additionally taking into account that a spin-1/2 is odd under time reversal $\Theta$~\cite{Bradley_Cracknell_1972,Wigner_1931}.

Roughly speaking, the coefficient functions $f_\mu(\bm{k})$ have to transform under the point-group operations like the corresponding matrices $\sigma_\mu$, i.e., they must be basis functions of the irreps or irreducible coreps corresponding to the $\sigma_\mu$. For monochromatic point groups, which are identical to the crystallographic point groups, the basis functions are found using representation theory \cite{Altmann_Herzig_1994, bilbao_server}. However, for gray point groups, which are characterized by the time-reversal operation $\Theta$ being a group element on its own, and for dichromatic point groups, for which $\Theta$ is not a group element but $\Theta$ multiplied by a unitary operation is, ordinary representation theory is not applicable due to the antiunitarity of $\Theta$.

There is a well-developed theory of coreps of magnetic groups \cite{Wigner_1931, Dimmock_Wheeler_1962, Cracknell_Wong_1967,Cracknell_1965, Bradley_Davies_1968,Bradley_Cracknell_1972}. However, this theory is dealing with \emph{complex} coreps, according to which operators and basis functions that are even or odd under time reversal can be equivalent. Since we need to distinguish these cases we have to use \emph{real} coreps. Note that we only have to consider single-valued coreps, not double-valued (spinor) coreps, because the Hamiltonian is a matrix of coefficient of fermionic bilinears.
In Appendix \ref{app:coreps}, we demonstrate how the character tables for gray and dichromatic groups are calculated using real corepresentation theory. Partially, the character tables for real coreps are also found in Ref.~\cite{Erb_Hlinka_2020}.

The eigenenergies of the general two-band Hamiltonian in Eq.\ \eqref{eq:general_twoband_Hamiltonian} are $E^\pm (\bm{k}) =f_0(\bm{k})\pm \sqrt{\sum_i f_i(\bm{k})^2}$. Hence, the valence and conduction bands, with energies $E^-$ and $E^+$, respectively, touch if 
\begin{equation}\label{eq:band_touchings_general}
	f_i(\bm{k})=0,\quad i=x,y,z. 
\end{equation}
Note that the coefficient function $f_0$ that corresponds to the identity matrix is not relevant for the band touchings. In three spatial dimensions, each equation $f_i(\bm{k})=0$ describes a two-dimensional surface so that the solutions of the system of equations \eqref{eq:band_touchings_general} are given by the intersections of three surfaces. In the absence of further constraints, these intersections are (possibly empty) sets of isolated points. Higher-dimensional intersections require that the number of independent equations in the system of equations \eqref{eq:band_touchings_general} is reduced by a symmetry such as a mirror symmetry. Besides isolated points, we only find one-dimensional lines of touching points.

In this paper, we expand the coefficient functions $f_\mu$ up to second order in momentum, or up to the first nonvanishing order if there are no contributions up to second order. This limits the maximal number of generic solutions given by isolated points of the system of equations \eqref{eq:band_touchings_general} according to Bézout's theorem \cite{Hulek2012}, which states that if each equation is polynomial, then the maximal number of generic solutions given by isolated points equals the product of the degrees of the three polynomials. 

In the final step, we investigate the robustness of one-dimensional solutions (nodal lines) by taking into account higher-order terms. This is necessary because the robustness of one-dimensional solutions of the system of equations \eqref{eq:band_touchings_general} is not guaranteed, in contrast to the robustness of zero-dimensional solutions, i.e., Weyl points, which are robust against symmetry-preserving perturbations as mentioned earlier.  

\subsection{Example}

Next, we illustrate the method by the example of the gray point group $\bm{M}=C_{3h}\otimes\{e,\Theta\}$ close to the $\Gamma$ point. Additional examples are given in Appendix \ref{app.examples}. First, let us give an intuitive explanation for the character table of $\bm{M}$ where we stick to the notation of Ref.~\cite{Altmann_Herzig_1994}. A more rigorous explanation is given in Appendix \ref{app.corep.gray}. The group $C_{3h}$ can be understood as $C_{3h}=C_3\otimes C_{1h}$ with $C_{1h}=\{e,\sigma_h\}$, where $\sigma_h$ denotes the mirror reflection in the horizontal plane. The real irreps of $C_3$ are $A$ and $E$. Since $\sigma_h$ commutes with all group elements of $C_3$, $C_{3h}$ has twice as many classes and real irreps as $C_3$. The real irreps of $C_{3h}$ are obtained from the real irreps of $C_3$ by labeling the latter with a prime if they are even under $\sigma_h$, and by a double prime if they are odd. The character table of $\bm{M}$ is constructed analogously, where the real irreducible coreps that are even and odd under $\Theta$ are labeled by the superscripts $+$ and $-$, respectively. Thus, $\bm{M}$ has the real irreducible coreps $A^{\prime+}$, $A^{\prime\prime+}$, $E^{\prime+}$, $E^{\prime\prime+}$, $A^{\prime-}$, $A^{\prime\prime-}$, $E^{\prime-}$, and $E^{\prime\prime-}$, as shown in Table \ref{table:char_tab_C3h_gray}. The pseudospin components $\sigma_i$ are odd under time reversal $\Theta$ and examination of their transformations under $C_{3h}$ shows that $\sigma_x$ and $\sigma_y$ transform according to $E^{\prime\prime-}$, while $\sigma_z$ transforms according to $A^{\prime-}$. The functions $f_i(\bm{k})$ must transform like the $\sigma_i$. The leading-order basis functions can be found in tables \cite{Erb_Hlinka_2020} or constructed by standard methods. $A^{\prime-}$ has the two leading-order basis functions $k_x(3k_y^2-k_x^2)$ and $k_y(3k_x^2-k_y^2)$ and Eq.\ \eqref{eq:band_touchings_general} becomes
\begin{equation}
f_z(\bm{k}) = a\, k_x(3k_y^2-k_x^2)+b\, k_y(3k_x^2-k_y^2) = 0 ,
\label{eq:coeff_c3hg_z}
\end{equation}
where $a,b\in \mathbb{R}$ are independent coefficients.

\begin{table*}
	\begin{center}
		\caption{Character table of the gray point group $C_{3h}\otimes\{e,\Theta\}$ with basis functions up to second or first nonvanishing order in momentum, and spin basis functions.}
		\begin{tabular}{c c  c  c   c c  c  c   c c}
			\hline\hline
			$C_{3h}\otimes\{e,\Theta\}$ & $e$  & $2C_3$ & $\sigma_h$ & $2S_3$ & $\Theta$  & $2\Theta C_3$ & $\Theta \sigma_h$ & $2\Theta S_3$ &  basis functions\\\hline
			\rule{0pt}{3ex}
			$A'^+$ & $1$ & $1$ & $1$ & $1$ & $1$ & $1$ & $1$ & $1$ & $\sigma_0, 1, k_x^2+k_y^2,k_z^2$\\
			\rule{0pt}{3ex}
			$A'^-$ & $1$ & $1$ & $1$ & $1$ & $-1$ & $-1$ & $-1$ & $-1$ & $\sigma_z, k_x(3k_y^2-k_x^2),k_y(3k_x^2-k_y^2)$\\
			\rule{0pt}{3ex}
			$A''^+$ & $1$ & $1$ & $-1$ & $-1$ & $1$ & $1$ & $-1$ & $-1$ & $ k_z k_x(3k_y^2-k_x^2), k_z k_y(3k_x^2-k_y^2)$ \\
			\rule{0pt}{3ex}
			$A''^-$ & $1$ & $1$ & $-1$ & $-1$ & $-1$ & $-1$ & $1$ & $1$ & $k_z$ \\
			\rule{0pt}{3ex}
			$E'^+$ & $2$ & $-1$ & $2$ & $-1$ & $2$ & $-1$ & $2$ & $-1$ & $\{k_x k_y,(k_x^2-k_y^2)/2\}$\\
			\rule{0pt}{3ex}
			$E'^-$ & $2$ & $-1$ & $2$ & $-1$ & $-2$ & $1$ & $-2$ & $1$ & $\{k_x,k_y\}$\\
			\rule{0pt}{3ex}
			$E''^+$ & $2$ & $-1$ & $-2$ & $1$ & $2$ & $-1$ & $-2$ & $1$ & $\{k_x k_z,k_y k_z\}$ \\
			\rule{0pt}{3ex}
			$E''^-$ & $2$ & $-1$ & $-2$ & $1$ & $-2$ & $1$ & $2$ & $-1$ & $\{\sigma_x,\sigma_y\},\{k_z k_x k_y, k_z (k_x^2-k_y^2)/2\}$\\
			\hline\hline
		\end{tabular}
		\label{table:char_tab_C3h_gray}
	\end{center} 	
\end{table*} 

The two-dimensional corep $E^{\prime\prime-}$ has the pair of basis functions $k_zk_xk_y$ and $k_z(k_x^2-k_y^2)/2$. There is an important point to note: For $\bm{M} = C_{3h}\otimes\{e,\Theta\}$, only the threefold rotation axis is uniquely determined, whereas all directions perpendicular to this axis are only determined up to an arbitrary rotation. Hence, it does not make sense to say which basis functions correspond to $\sigma_x$ and $\sigma_y$. This is reflected by the equation
\begin{equation}
\begin{pmatrix}
  f_x(\bm{k}) \\
  f_y(\bm{k})
\end{pmatrix} = c\,k_z \,R_\alpha
\begin{pmatrix}
  k_x k_y \\
  \frac{k_x^2-k_y^2}{2}
\end{pmatrix} = 0 , \label{eq:coeff_c3hg_x_y}
\end{equation}
where $c$ is a coefficient independent of $a$ and $b$, and $R_\alpha$ is a two-dimensional rotation matrix with the arbitrary rotation angle $\alpha$.

\begin{figure}
	\begin{center}
		\includegraphics[width=\columnwidth]{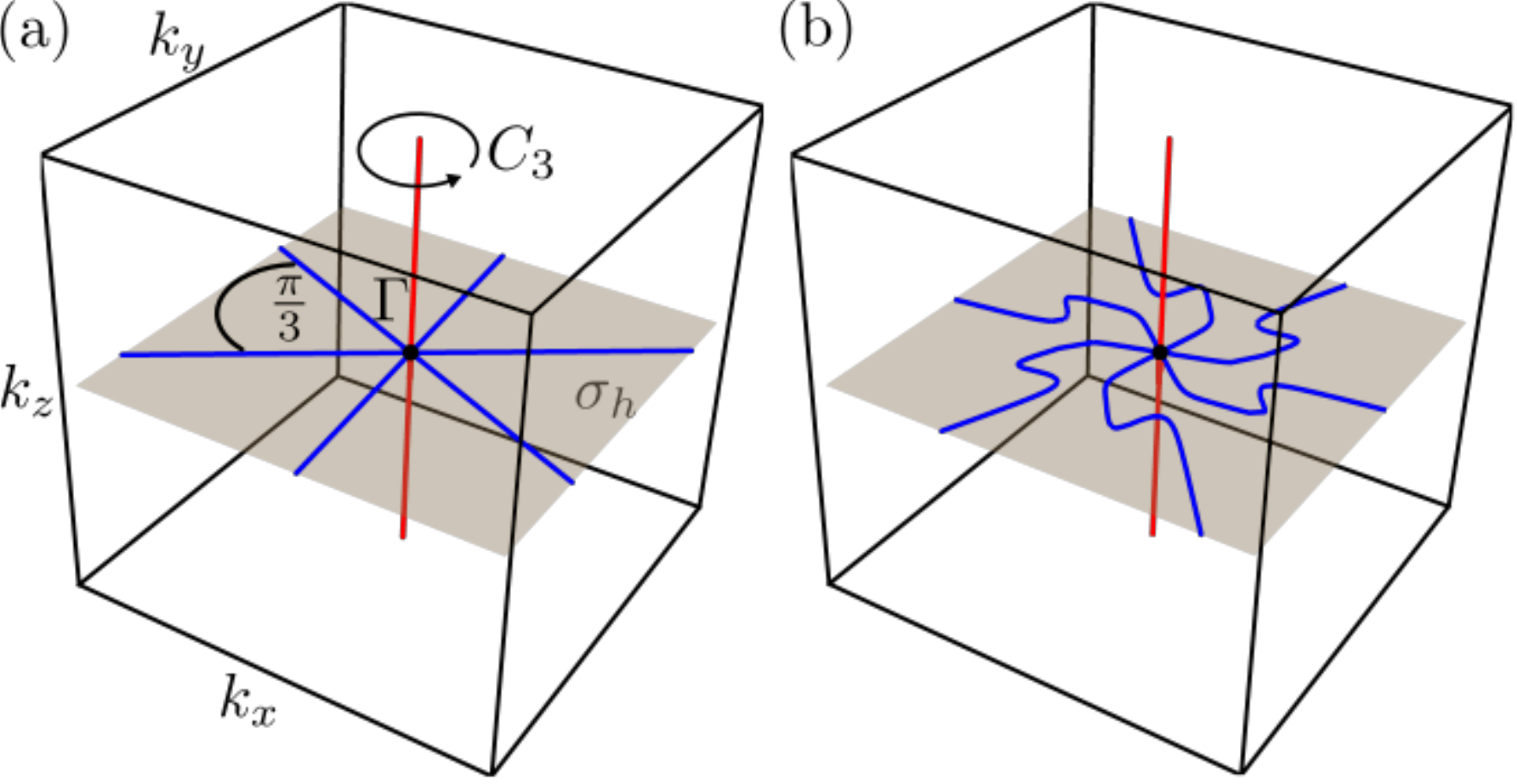}
	\end{center}
	\caption{Band touchings of the gray point group $C_{3h}\otimes\{e,\Theta\}$. (a) For terms up to third order in momentum, the band touchings are given by one nodal line on the threefold rotation axis (red) and three nodal lines (blue) in the mirror plane (gray) intersecting each other at the $\Gamma$ point. (b)  When higher-order terms are taken into account, the nodal line on the $k_z$ axis remains unchanged, whereas the nodal lines in the mirror plane are deformed but retain their sixfold symmetry.}\label{fig_1}
\end{figure}

One solution of Eqs.~\eqref{eq:coeff_c3hg_z} and \eqref{eq:coeff_c3hg_x_y} is given by $k_x=k_y=0$, i.e., by the $k_z$ axis. Another solution is found in the $k_x k_y$ plane. Setting $k_z=0$ solves Eq.\ \eqref{eq:coeff_c3hg_x_y}. By introducing polar coordinates $k_x = \rho \cos \phi$ and $k_y = \rho \sin \phi$, Eq.\ \eqref{eq:coeff_c3hg_z} becomes
\begin{equation}
  f_z(\rho,\phi,k_z=0) = \rho^3\, (- a \cos 3\phi +b \sin 3\phi) = 0.
\label{eq:coeff_c3hg_z_polar}
\end{equation}
The solutions that are not already contained in the solution with $k_x=k_y=0$ are
\begin{equation}
  \phi = \frac{1}{3}\, \arctan\frac{a}{b} + \frac{\pi}{3}\, n ,
\end{equation}  
with $n\in\mathbb{Z}$. Hence, there are six solutions for $\phi$ in the interval $[0,2\pi)$ that differ by multiples of $\pi/3$. As depicted in Fig.~\ref{fig_1}(a), there are in total four nodal lines: one nodal line on the $k_z$ axis, which is the threefold rotation axis, and three straight nodal lines with sixfold symmetry in the $k_x k_y$ plane, which is the mirror plane. Note that the sixfold symmetry is a consequence of the presence of $\Theta$ since time reversal acts on the momentum $\bm{k}$ by inverting its sign, $\hat{P}_\Theta \bm{k}=-\bm{k}$, similar to inversion symmetry. Together with the threefold rotation symmetry in $C_{3h}$, this yields a sixfold symmetry. 

\begin{table*}
	\caption{Band touchings for the 32 gray and the 32 monochromatic point groups. Weyl points can occur on and off rotation axes, where integers specify the possible numbers of Weyl points. Weyl points at the reference point (for example, the $\Gamma$ point) are denoted by RP. Nodal lines are found on rotation axes or in mirror planes (MPs). Open nodal lines on axis are denoted by $n C_m$, where $n$ specifies the number of nodal lines on the rotation axes, which have $m$-fold rotation symmetry and might be further distinguished by primes. In MPs, there are closed (C) and open nodal lines. The latter are denoted by $k\sigma_h$, $k\sigma_v$, and $k\sigma_d$ for nodal lines in horizontal, vertical, and diagonal mirror planes, respectively, where $k$ is the number of lines. Nodal lines that are not mandatory are indicated by \noline. ``$+$'' indicates band touchings that occur simultaneously, whereas commas separate band touchings of which only one occurs in a particular model. If a band touching is not possible at all the entry is left blank.
	\label{table:mono_gray}}
	\begin{flushleft}	
		\begin{tabularx}{\textwidth}{P{.03\textwidth}P{.10\textwidth}P{.12\textwidth}P{.14\textwidth}P{.10\textwidth}P{.1\textwidth}P{.10\textwidth}P{.14\textwidth}P{.1\textwidth}}
			
			\hline \hline
			\# & Point group  & \multicolumn{3}{c}{Gray point group} &  \multicolumn{4}{c}{Monochromatic point group}\\ 
			\cmidrule(lr){3-5} \cmidrule(lr){6-9}
			& &  \multicolumn{1}{c}{Weyl points} & \multicolumn{2}{c}{Nodal lines} & \multicolumn{2}{c}{Weyl points} & \multicolumn{2}{c}{Nodal lines} \\
			\cmidrule(lr){3-3} \cmidrule(lr){4-5} \cmidrule(lr){6-7} \cmidrule(lr){8-9}
			& &  \multicolumn{1}{c}{On axis} &  \multicolumn{1}{c}{On axis} & \multicolumn{1}{c}{In MP} & \multicolumn{1}{c}{On axis}& \multicolumn{1}{c}{Off axis} & \multicolumn{1}{c}{On axis} & \multicolumn{1}{c}{In MP} \\ \hline
			
			1 & $C_1$ & RP & & &  & $0,\, 2,\, 4,\, 6,\, 8$ &   &     \\ 
			
			2 & $C_2$ & RP &  & & $0,\, 2$ & $0,\, 2,\, 4$ &  &    \\				
			
			3 & $C_3$ & RP &  & & $0,\, 2$ & $0,\, 6$ &  &     \\		
			
			4 & $C_4$ & RP &  & & $0,\, 2$ &  &  &    \\	
			
			5 & $C_6$ & RP &  & & $0,\, 2$ &  &  &   \\
			
			6 & $D_2$ & RP &  & & RP & $0,\, 4$ &  &    \\	
			
			7 & $D_3$  & RP &   & & RP + $3$ &  &  &    \\		
			
			8 & $D_4$  & RP &   &  & RP &  &  &   \\	
			
			9 & $D_6$ & RP &  & & RP &  &  &     \\
			
			10 & $T$ & RP &  & & RP + 4 &  &  &     \\ 
			
			11 & $O$ & RP &   & & RP &  &  &    \\	
			
			12 & $C_i$ &  &  & &  & $0,\, 4,\, 8$ & &     \\ 
			
			13 & $S_4$ &  & $1\,C_2$ & & $0,\, 2$ & $0,\, 4$ & &    \\	 
			
			14 & $S_6$ &  &  & & $0,\, 2$ &  $0,\, 6$ & &     \\

			15 & $C_{1h}$ &  &  & $1\,\sigma_h$ &  & $0,\, 2$ &  & $0,\, 2\,\sigma_h,$ C   \\				
			
			16 & $C_{2h}$ &  &  & & $0,\, 2$ &  &  & $0,\,2\,\sigma_h,$ C   \\	
			
			17 & $C_{3h}$  &  & $1\,C_3$  & $3\,\sigma_h$ & $0,\, 2$ &  &  & \noline, C  \\
			
			18 & $C_{4h}$ &  &  & & $0,\, 2$ &  &  & \noline, C    \\
			
			19 & $C_{6h}$  & &   & & $0,\,2$ & &  & \noline, C   \\

			20 & $C_{2v}$ &  & $1\,C_2$  & &  &  & $1\, C_2$ & $2\,\sigma_v$     \\	
			
			21 & $C_{3v}$ &  & $1\,C_3$  &  &  &  & $1\,C_3$ & $3\,\sigma_v$   \\			
			
			22 & $C_{4v}$ &  & $1\,C_4$  & &  &  & $1\, C_4$ & $4\, \sigma_v$    \\		
			
			23 & $C_{6v}$ &  & $1\,C_6$  & &  &  & $1\, C_6$ & $6\,\sigma_v$  \\	
			
			24 & $D_{2d}$ &  & $1\,C_2$  & &  &  & $1\, C_2$ & $2\, \sigma_d$    \\
			
			25 & $D_{3d}$ &  &   & &  &  & $1\,C_3$ & $3\,\sigma_d$   \\	
			
			26 & $D_{2h}$ &  &  & &  &  & $3\,C_2$ &    \\	
			
			27 & $D_{3h}$ &  & $1\,C_3+3\,C_2^\prime$  & &  &  & $1\,C_3+3\,C_2^\prime$ &     \\
			
			28 & $D_{4h}$ &  &   & &  &  & $1\,C_4+2\, C_2^\prime+2\, C_2^{\prime\prime}$ &     \\
			
			29 & $D_{6h}$ &  &   & &  &  & $1\,C_6+3\,C_2^\prime+3\,C_2^{\prime\prime}$ &    \\
			
			30 & $T_h$ & & & &  &  & $3\,C_2$ &   \\
			
			31 & $T_d$ &  & $3\,C_2+4\,C_3$  & &  &  & $3\,C_2+4\,C_3$ &     \\	
			
			32 & $O_h$  &  &   &  &  &  & $3\,C_4+4\,C_3+6\,C_2$ &  \\	
			
			\hline \hline
		\end{tabularx}	
	\end{flushleft}
\end{table*}

Next, we investigate the robustness of the nodal lines by taking into account higher-order terms for the coefficient functions. First, we note that the relevant coreps $A^{\prime-}$ and $E^{\prime\prime-}$ do not have any basis functions that are pure powers of $k_z$ (these belong to $A^{\prime+}$ or $A^{\prime\prime-}$). Hence, one solution of Eq.\ \eqref{eq:band_touchings_general} up to arbitrary order is always given by $k_x=k_y=0$. Consequently, the nodal line on the threefold rotation axis is unaffected by higher orders as depicted by the red line in Fig.~\ref{fig_1}(b). We call a nodal line that is left unchanged by higher-order terms a nodal line ``on axis.''

For the nodal lines in the $k_x k_y$ plane, we use that the momentum basis functions of the two-dimensional $E''^-$ corep are odd in $k_z$. Hence, one solution of $f_x=f_y=0$ is always given by $k_z=0$. Thus, the three nodal lines stay in the mirror plane. Inclusion of higher-order terms renders Eq.\ \eqref{eq:coeff_c3hg_z} complicated but does not prevent solutions for a discrete sets of angles $\phi$, which now depends on the coordinate $\rho$. Hence, these nodal lines are not destroyed but deformed. This is illustrated by the blue lines in Fig.~\ref{fig_1}(b). The nodal lines that are deformed but not destroyed by higher-order terms are called ``deformable'' nodal lines. It is natural that the deformable nodal lines do not remain straight beyond leading order since $\bm{M}$ does not contain any symmetry that requires them to be straight. Note that the sixfold symmetry survives.

Both nodal lines on axis and deformable nodal lines are robust in the sense that they are not destroyed by higher-order terms. A nodal line on axis can only exist on a rotation axis, whereas a deformable nodal line can only exist in a mirror plane. As we shall see, the reverse statements do not hold: The existence of a rotation axis or a mirror plane does not guarantee the existence of a nodal line on axis or a deformable nodal line, respectively.

In addition to being deformable or on axis, a robust nodal line can also be open or closed. A closed nodal line is a one-dimensional compact manifold while an open nodal line is not compact. All nodal lines in our example are open. For the rest of this section, we use ``trivial irrep'' for both the trivial irrep and the trivial irreducible corep to simplify the discussion. Since our method focuses on analyzing band touchings in the vicinity of RPs without taking into account the periodicity of the Brillouin zone, a nodal line on axis is always open, while deformable nodal lines can be either open or closed depending on the group. Closed deformable nodal lines can exist if one of the three equations in Eq.~\eqref{eq:band_touchings_general} corresponds to a Pauli matrix that is in the trivial irrep, while the other two equations are solved on a two-dimensional manifold, i.e., a plane. Then, if we include only terms up to second order in momentum the equation corresponding to the Pauli matrix in the trivial irrep is given by a quadric
\begin{equation}\label{eq:quadric}
p + \bm{q}\cdot\bm{k} + \bm{k}^T Q\, \bm{k} = 0,
\end{equation} 
where $p\in\mathbb{R}$, $\bm{q}\in\mathbb{R}^3$, and $Q\in\mathbb{R}^{3\times3}$ include coefficients that are characteristic for the point group under consideration. Since for the trivial irrep $p$ is generically nonzero Eq.~\eqref{eq:quadric} can describe an ellipsoid, an elliptic or hyperbolic cylinder, parallel planes, a hyperboloid of one sheet or two sheets, or an empty set. Hence, depending on the type of the quadric, on the location of its center, and on the orientations of its principal axes with respect to the plane, there can be no intersection of the quadric and the plane, or there is an open or closed line. If the plane corresponds to a mirror plane, the nodal line is deformable. In particular, we expect that higher-order terms deform a deformable nodal line in the mirror plane but do not open (close) a closed (open) nodal line. At this point, we emphasize that only a closed deformable nodal line necessarily requires that one of the Pauli matrices belongs to the trivial irrep but not an open one.

\section{Results}

In a manner similar to the previous example, we have investigated the band touchings for the 122 magnetic point groups. We discuss gray, monochromatic, and dichromatic point groups in turn.

\subsection{Gray and monochromatic point groups}

We start by discussing the band touchings for the gray point groups, which are summarized in Table \ref{table:mono_gray}. For gray point groups, Weyl points can appear on rotation axes, whereas nodal lines can occur on rotation axes and in mirror planes. We find that all gray chiral point groups (\#1--11), i.e., point groups that explicitly contain $\Theta$ but lack inversion, mirror symmetries, and improper rotations, give rise to a Weyl point at the RP. Gray point groups can only occur as little groups for RPs that correspond to time-reversal-invariant momenta so that our results are in accordance with the findings of Chang \textit{et al.}\ \cite{Chang_Wieder_2018}, who state that so-called Kramers-Weyl fermions appear at time-reversal-invariant momenta in nonmagnetic chiral crystals with spin-orbit coupling. Note that Weyl points at time-reversal-invariant momenta should lead to long Fermi arcs in surface Brillouin zones, which are expected to be different for the two chiral enantiomers. Quasiparticle interference has provided experimental evidence for such Fermi arcs in PdGa (gray point group $T\otimes\{e,\Theta\}$)~\cite{Sessi2020}.

Next, we turn our attention to the achiral gray point groups (\#12--32). Gray point groups with inversion symmetry (\#12, 14, 16, 18, 19, 25, 26, 28, 29, 30, 32) contain the product $\Theta i$. The Kramers theorem \cite{Kramers_1930} then ensures that the two bands are degenerate at all $\bm{k}$. Hence, there are no Weyl points or nodal lines. The remaining achiral gray point groups that lack inversion symmetry are guaranteed to exhibit open nodal lines. All gray point groups that have at least one rotation axis that is located within at least one mirror plane (\#20, 21, 22, 23, 24, 27, 31) give rise to open nodal lines along all rotation axes that are located in at least one mirror plane. The remaining three achiral gray point groups left to discuss are $S_4\otimes\{e,\Theta\}$, $C_{1h}\otimes\{e,\Theta\}$, and $C_{3h}\otimes\{e,\Theta\}$. $S_4\otimes\{e,\Theta\}$ is the only gray point group that possesses an improper rotation axis, which is also a twofold rotation axis, but lacks inversion and mirror symmetry. In this case, we still find an open nodal line on the rotation axis, which is in accordance with Ref.~\cite{Bouhon_Lange_2021}. $C_{1h}\otimes\{e,\Theta\}$ lacks a rotation axis but still exhibits an open nodal line in its mirror plane. $C_{3h}\otimes\{e,\Theta\}$ has an open nodal line in its mirror plane similar to $C_{1h}\otimes\{e,\Theta\}$, and possesses an open nodal line on its rotation axis, which is also an improper rotation axis, similar to $S_4\otimes\{e,\Theta\}$. Our findings for the achiral gray point groups agree with the results of Xie \textit{et al.}\ \cite{Xie_Gao_21}. The authors report so-called Kramers nodal lines, which are protected by a combination of time-reversal symmetry and achiral symmetries, in achiral noncentrosymmetric materials with spin-orbit coupling \cite{Xie_Gao_21}. Kramers nodal lines are doubly degenerate band-touching lines that connect time-reversal invariant momenta. In our analysis, all achiral gray point groups that lack inversion symmetry host at least one guaranteed open nodal line, which is consistent with Kramers nodal lines connecting time-reversal invariant momenta.

We now turn to the band touchings for the monochromatic point groups that are also summarized in Table \ref{table:mono_gray}. Monochromatic point groups can host Weyl points on and off rotation axes and nodal lines on rotation axes and in mirror planes. To begin with, we discuss the Weyl points on axis for chiral monochromatic point groups (\#1--11). In contrast to the chiral gray point groups, not all chiral monochromatic point groups possess Weyl points at the RP. Specifically, cyclic chiral monochromatic point groups with at most one proper rotation axis, i.e., $C_n$ ($n=1,2,3,4,6$), do not possess a Weyl point at the RP. The reason for this is that for these point groups $\sigma_z$ belongs to the trivial irrep. Consequently, the coefficient function $f_z$ contains a constant term so that the system of equations \eqref{eq:band_touchings_general} generically cannot be solved at the RP, as also predicted in Ref.~\cite{Fang_Gilbert_2012}. For the noncyclic chiral monochromatic point groups $D_2$, $D_3$, $D_4$, $D_6$, $T$, and $O$, which have more than one rotation axis, a Weyl point at the RP is guaranteed. The reason is that for these groups no pseudospin component belongs to the trivial irrep. Hence, the RP ($\bm{k}=0$) always solves Eq.~\eqref{eq:band_touchings_general}. Furthermore, our analysis shows that $D_3$ and $T$ have in addition to the Weyl point at the RP three and four Weyl points on rotation axes in the vicinity of the RP, respectively. In Appendix \ref{app.examples_T}, we show how the band touchings for the point group $T$ are calculated. Note that noncyclic monochromatic point groups can only appear as little groups of high-symmetry points but not for momenta on high-symmetry axes.  Otherwise, there would be a ``Weyl point'' at every point of a high-symmetry axis, and these points would form a nodal line, in contradiction to the absence of a nodal line on axis in the analysis of the high-symmetry points. While we have already mentioned in the discussion of gray point groups that nonmagnetic chiral materials have Kramers-Weyl points at all time-reversal-invariant momenta, such a statement does not hold for materials that belong to monochromatic point groups. For example, the time-reversal invariant momentum at $(\frac{1}{2}\frac{1}{2}\frac{1}{2})$ for the orthorombic space group $F222$ has the little group $C_1$ \cite{bilbao_server} and hence no Weyl point.

Next, we turn our attention to the achiral monochromatic point groups (\#12--32). With respect to their band touchings, these groups can be divided into four types. The first type comprises the monochromatic point groups that lack mirror symmetries, i.e., $C_i$, $S_4$, and $S_6$. Groups of the first type can host Weyl points on and off axis but no nodal lines due to the lack of mirror symmetry. To the second type belong groups with a single mirror plane, i.e., $C_{nh}$ ($n=1,2,3,4,6$). For groups of the second type, band-touching points and lines are not guaranteed to exists since one of the Pauli matrices is found in the trivial irrep. They can host a pair of two Weyl points and a closed nodal line in the mirror plane. For $C_{1h}$ and $C_{2h}$, the nodal line can be open instead of closed. Achiral monochromatic point groups of the first and second type have in common that band touchings are not guaranteed. 

Guaranteed band touchings are characteristic for achiral monochromatic point groups of the third and fourth type. The monochromatic point groups belonging to the third type are characterized by one rotation axis common to all mirror planes (\#20--25). These groups host open nodal lines in all of their mirror planes and along their common axis. To the fourth type belong groups that have multiple rotation axes located in at least one mirror plane (\#26--32). For these groups, all axes that are located in mirror planes host open nodal lines.  

\begin{table*}
	\caption{Band touchings for the 37 dichromatic point groups $\bm{G}(\bm{H})$ that do not contain the product of inversion and time-reversal symmetry. Weyl points can occur on and off rotation axes, where integers specify the possible numbers of Weyl points. Weyl points at the reference point (for example, the $\Gamma$ point) are denoted by RP. Robust nodal lines are found on rotation axes or in mirror planes (MPs), whereas fragile nodal lines, which are not robust against higher-order terms, are found in a plane perpendicular to a rotation axis. Open nodal lines on axis are denoted by $n C_m$, where $n$ specifies the number of nodal lines on the rotation axes with $m$-fold rotation symmetry. In MPs, there are closed (C) and open nodal lines. The latter are denoted by $k\sigma_h$, $k\sigma_v$, and $k\sigma_d$ for nodal lines in horizontal, vertical, and diagonal mirror planes, respectively, where $k$ is the number of lines. Nodal lines that are not mandatory are indicated by \noline. ``$+$'' indicates band touchings that occur simultaneously, whereas commas separate band touchings of which only one occurs in a particular model. If a band touching is not possible at all the entry is left blank.
		\label{table:di}}
	\begin{flushleft}	
		\begin{tabularx}{\textwidth}{P{.03\textwidth}P{.10\textwidth}P{.16\textwidth}P{.17\textwidth}P{.17\textwidth}P{.16\textwidth}P{.16\textwidth}}
			
			\hline \hline
			\# & Point group  & \multicolumn{2}{c}{Weyl points} &   \multicolumn{3}{c}{Nodal lines}\\ 
			\cmidrule(lr){3-4} \cmidrule(lr){5-7}
			& &  On axis & Off axis & On axis & In MP & Fragile \\ \hline
			
			1& $C_2(C_1)$  & & $0,\,2,\,4,\,6,\,8$   & &  &  \\ 
			
			2&$C_{4}(C_{2})$ & RP & & & & \\
			
			3&$C_{6}(C_{3})$ & RP & $3$ & &  &  \\
			
			4&$D_{2}(C_2)$ & $0,\,2$ & $0,\,2,\,4$  & & & \\
			
			5&$D_{3}(C_{3})$ & $0,\,2$ &   $0,\,6$ & &  & \\ 
			
			6&$D_{4}(C_{4})$ & $0,\,2$ &  & & & \noline, C \\
			
			7&$D_{6}(C_{6})$ & $0,\,2$ &  & &  & \noline, C \\ 
			
			8&$D_{4}(D_{2})$ & RP &  $0,\,4$ & & & \\
			
			9&$D_{6}(D_{3})$ & RP + 3 &   &  &  &  \\ 			
			
			10&$O(T)$ & RP + 4 &   & & &  \\
			
			11&$C_{2v}(C_{2})$ & $0,\,2$ & $0,\,4$   & & & \\
			
			12&$C_{3v}(C_{3})$ & $0,\,2$ & $0,\,6$  & &  &  \\
			
			13&$C_{4v}(C_{4})$ & $0,\,2$ & &  & &  \\
			
			14&$C_{6v}(C_{6})$ & $0,\,2$ &   & & &  \\
			
			15&$C_{1h}(C_1)$ & & $0,\,2,\,4,\,6,\,8$  &  & & \\
			
			16&$C_{3h}(C_{3})$ & &   & $1\,C_3$ &  &  \\
			
			17&$S_{4}(C_{2})$ & &  & $1\,C_2$ & & \\
			
			18&$D_{2d}(D_{2})$ & &  & $1\,C_2$ (with $\Theta S_4$ sym.) & &  \\
			
			19&$D_{3h}(D_{3})$ & &   & $1\,C_3$ & &  \\
			
			20&$T_d(T)$ &  &   & $3\,C_2$ &  &  \\
			
			21&$C_{2h}(C_i)$ & & $0,\,4,\,8$ &  & & \\
			
			22&$D_{2d}(S_{4})$ & $0,\,2$ & $0,\,4$  &  & &  \\
			
			23&$D_{3d}(S_{6})$ & $0,\,2$ &  $0,\,6$ & &  &  \\
			
			24&$C_{6h}(S_{6})$ & &  &  $1\, C_3$ &  &  \\
			
			25&$C_{2v}(C_{1h})$ &  & $0,\,2,\,4$  & & \noline, $2\,\sigma_h$, C & \\
			
			26&$D_{2h}(C_{2h})$ & $0,\,2$ & $0,\,2,\,4$   & & \noline, $2\,\sigma_h$, C & \\
			
			27&$D_{3h}(C_{3h})$ & $0,\,2$ &  &  & \noline, C &  \\
			
			28&$D_{4h}(C_{4h})$ & $0,\,2$ & &  &  \noline, C &  \\
			
			29&$D_{6h}(C_{6h})$ & $0,\,2$ &  &  & \noline, C &  \\
			
			30&$C_{4h}(C_{2h})$ &  &  & $1\,C_2$ &  $2\,\sigma_h$ & \\
			
			31&$C_{4v}(C_{2v})$ & &  & $1\,C_2$ &  $2\,\sigma_v$ &  \\
			
			32&$C_{6v}(C_{3v})$ &    & & $1\,C_3$ &  $3\,\sigma_v$ &  \\
			
			33&$D_{2d}(C_{2v})$ &  & & $1\,C_2$ &  $2\,\sigma_v$ &  \\
			
			34&$D_{3h}(C_{3v})$ &   & & $1\,C_3$ & &  \\
			
			35&$D_{4h}(D_{2h})$ &   & & $3\,C_2$ & &  \\
			
			36&$D_{6h}(D_{3d})$ &    & & $1\,C_3$ &  &  \\
			
			37&$O_h(T_h)$ &  &  &  $3\,C_2$ &  &  \\					
			\hline \hline
		\end{tabularx}
	\end{flushleft}
\end{table*}

\subsection{Dichromatic point groups}

For dichromatic point groups that contain the product of inversion and time-reversal symmetry, the two bands are degenerate for all $\bm{k}$ \cite{Kramers_1930}, as discussed above. It is thus sufficient to focus on the band touchings for the 37 dichromatic point groups that do not contain this product. These groups are listed in Table \ref{table:di}. We denote every dichromatic point group in the form $\bm{G}(\bm{H})$ where $\bm{H}$ is a halving subgroup of $\bm{G}$ that contains all unitary elements of $\bm{G}(\bm{H})$, whereas all antiunitary elements of $\bm{G}(\bm{H})$ are given by the set $\Theta (\bm{G}-\bm{H})$ \cite{Dresselhaus_2007}. For dichromatic point groups, there can be Weyl points on and off rotation axes and robust nodal lines on axis and in mirror planes, as well as fragile line nodes, which are described below in more detail.

First, we discuss chiral dichromatic point groups $\bm{G}(\bm{H})$, for which both $\bm{G}$ and $\bm{H}$ are chiral groups (\#1--10). A chiral dichromatic point group $\bm{G}(\bm{H})$ hosts a Weyl point at the RP if both $\bm{G}$ and $\bm{H}$ have only one rotation axis [$C_4(C_2)$, $C_6(C_3)$] or if both have multiple rotation axes [$D_4(D_2)$, $D_6(D_3)$, $O(T)$]. Furthermore, $C_6(C_3)$ is the only magnetic point group that is guaranteed to host a triple of Weyl points off axis. $C_6$ and $C_3$ are subgroups of $D_6$ and $D_3$, respectively. For $D_6(D_3)$, there are one Weyl point at the RP and three guaranteed Weyl points on the three twofold rotation axes. Thus, we can understand $C_6(C_3)$ in terms of $D_6(D_3)$, where $C_3$ lacks the twofold rotation axes that are present in $D_3$ and for that group pin the off-axis Weyl points to the rotation axes. Due to the lack of any achiral symmetries that could protect nodal lines, chiral dichromatic point groups do not host any robust nodal lines. However, applying the method of Sec.\ \ref{sec:method} to $D_4(C_4)$ and $D_6(C_6)$, we find that closed nodal lines are possible. Since neither $D_4(C_4)$ nor $D_6(C_6)$ have any achiral symmetries that could protect a nodal line these lines are \emph{fragile}, i.e., they are not robust against higher-order terms. When terms of sufficiently high order are taken into account, fragile nodal lines are split into Weyl points as shown in more detail for $D_4(C_4)$ in Appendix~\ref{app.example.D4C4}.

Next, we focus on semichiral dichromatic point groups $\bm{G}(\bm{H})$, for which $\bm{G}$ is an achiral group and $\bm{H}$ is a chiral group (\#11--20). We find that semichiral point groups can host Weyl points (\#11--15) or nodal lines on axis (\#16--20). Nodal lines on axis require the product of time reversal $\Theta$ and an improper rotation axis. We have already seen for the gray point group $S_4\otimes\{e,\Theta\}$ that $\Theta S_4$ can protect nodal lines on axis.

Finally, we discuss achiral dichromatic point groups $\bm{G}(\bm{H})$, for which both $\bm{G}$ and $\bm{H}$ are achiral point groups (\#21--37). There are four achiral dichromatic point groups that lack any mirror symmetry (\#21--24). Three of these groups,  $C_{2h}(C_i)$, $D_{2d}(S_4)$, and $D_{3d}(S_6)$, can host only Weyl points but no nodal lines. The fourth of them, the group $C_{6h}(S_6)$, has a nodal line on its threefold rotation axis, which has also $\Theta S_3$ symmetry. The next five groups (\#25--29) have in common that they can host Weyl points and nodal lines in mirror planes. This is due to the fact that for $C_{2v}(C_{1h})$ and for $D_{nh}(C_{nh})$ with $n=2,3,4,6$ one Pauli matrix is in the trivial irreducible corep. Hence, their band touchings are similar (but not necessarily identical) to the band touchings for the monochromatic point groups $C_{nh}$. For the remaining achiral dichromatic point groups, we find that they either host nodal lines on axis and in their mirror planes (\#30--33) or that they host only nodal lines on axis (\#34--37). Note that all dichromatic point groups with the halving subgroup $C_{2v}$ or $C_{3v}$ have guaranteed open nodal lines. Furthermore, we remark that the groups $C_{4h}(C_{2h})$, $D_{3h}(C_{3v})$, $D_{4h}(D_{2h})$, $D_{6h}(D_{3d})$, and $O_h(T_h)$, which have at least one rotation axis with $\Theta S_n$ symmetry, give rise to open nodal lines on axis.   

To the best of our knowledge, there is no study that classifies the band touchings for all dichromatic point groups that lack the product of time-reversal and inversion, for which reason we compare our results in Table \ref{table:di} to two case-by-case studies. The first work by Wang \textit{et al.}\ \cite{Wang_Zhao_2018} is based on first-principles calculations and reports a nodal ring in the mirror plane in the vicinity of the $\Gamma$ point for magnetic oxides with the dichromatic point group $D_{4h}(C_{4h})$. This agrees with our results but the authors have overlooked the possibility of a pair of Weyl points on the $k_z$ axis. Second, Jin \textit{et al.}\ \cite{Jin_Wang_2017} investigate band touchings close to the $\Gamma$ point in the $k_x k_y$ plane in tetragonal structures with ferromagnetic ordering as found in $\beta$-V$_2$OPO$_4$, Co$_2$S$_2$Tl, and Fe$_2$S$_2X$ ($X = \mathrm{Al}, \mathrm{Ga}, \mathrm{In}$), where different magnetization directions are considered. If the magnetization direction is along the $[001]$ direction, the little group of $\Gamma$ is again given by $D_{4h} (C_{4h})$, and a nodal ring is observed just as in Ref.\ \cite{Wang_Zhao_2018}. For spins directed along the $[110]$ direction, the symmetry is lowered to $D_{2h}(C_{2h})$ and the ring node splits into a pair of Weyl points that are separated along the $\left[110\right]$ axis, which is again consistent with our results in Table \ref{table:di}.

\subsection{Off-axis Weyl points and lattice models}

So far, we have omitted discussing off-axis Weyl points. Such Weyl points can appear for monochromatic and dichromatic point groups, as reflected by Tables \ref{table:mono_gray} and \ref{table:di}, respectively. Since we investigate the band touchings in the vicinity of a RP by expanding a pseudospin-1/2 Bloch Hamiltonian up to second or the first nonvanishing order in momentum the total number of Weyl points is limited by Bézout's theorem \cite{Hulek2012}, as mentioned earlier. Hence, our tables report the possible numbers of off-axis Weyl points that are compatible with the point group and with Bézout's theorem. However, there may be more Weyl points resulting from higher-order terms in the basis functions $f_i(\bm{k})$.

\begin{figure}[h]
	\raisebox{36ex}[0ex][2ex]{(a)}\hspace{0em}\includegraphics[width=0.63\columnwidth]{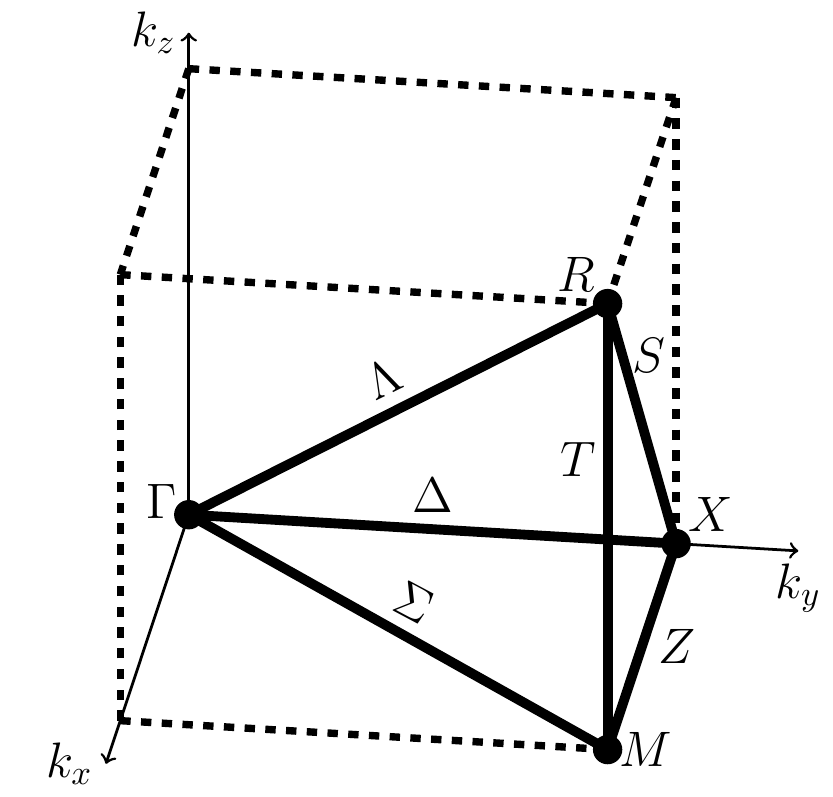}\\
	\raisebox{35ex}[0ex][2ex]{(b)}\hspace{0em}\includegraphics[width=0.63\columnwidth]{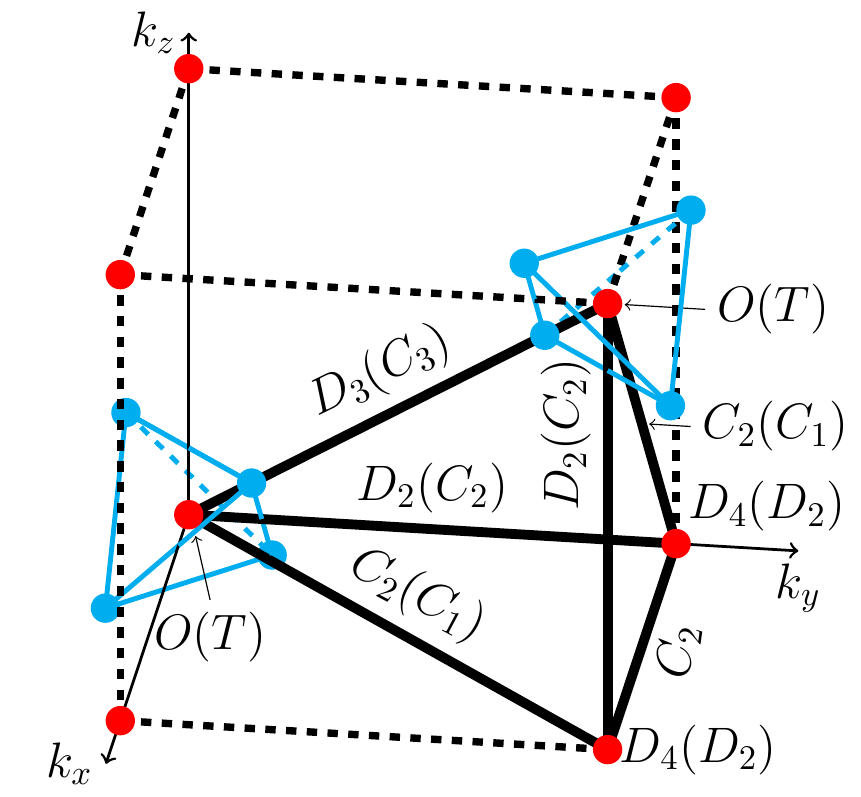}\\
	\raisebox{35.6ex}[0ex][0ex]{(c)}\hspace{0em}\includegraphics[width=0.63\columnwidth]{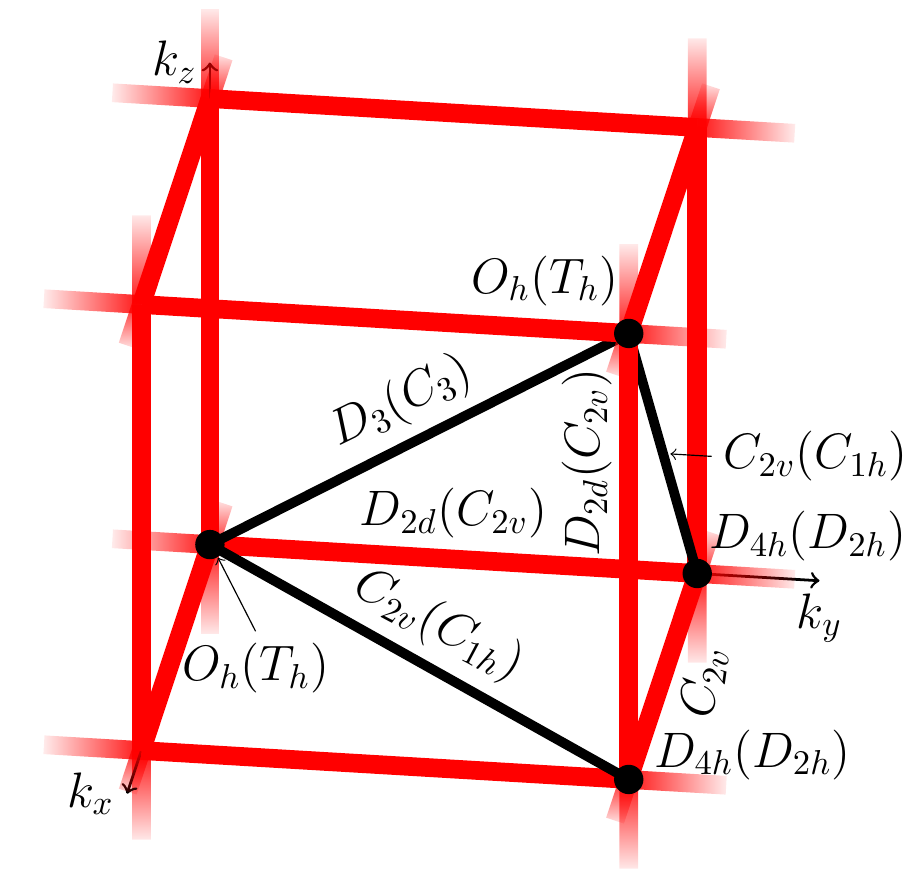}
	\caption{Weyl points and nodal lines for the two lattice models $A$ and $B$. (a) One octant of the simple cubic Brillouin zone with high-symmetry points and lines. (b) Band touchings for lattice model $A$. The high-symmetry points and lines are labeled by the corresponding little groups \cite{Elcoro_Wieder_2021, Xu_Elcoro_2020}. Weyl points that are fixed to high-symmetry points are displayed in red, while other Weyl points are displayed in cyan. (c) Band touchings for lattice model $B$. The labels are analogous to panel (b). Nodal lines are displayed in red. }\label{fig_2}
\end{figure}

It remains to discuss how our analysis for the vicinity of RPs transfers to lattice models. As noted above, for lattice models, the total chirality of all Weyl points in the Brillouin zone vanishes \cite{Armitage_Mele_2018,Nielsen_Ninomiya_1983}. However, this result does not apply to continuum models. Hence, odd numbers of Weyl points can and do appear in our analysis. Furthermore, for lattice models, the open nodal lines must be compatible with the periodicity of momentum space. For RPs in a Brillouin zone, our analysis shows if and where a nodal line can be found. The global structure of nodal lines on axes is then fixed. How open deformable nodal lines connect in periodic momentum space is not studied in the present work. However, we expect that the symmetry of the nodal system under the magnetic point group and the analysis of the vicinity of high-symmetry points in the mirror plane usually constrain the connectivity of these line nodes. In order to illustrate the application of our method to lattice models, we consider two models $A$ and $B$ with the dichromatic point groups $O(T)$ and $O_h(T_h)$, respectively. Both models assume a simple cubic lattice. Figure \ref{fig_2}(a) shows one octant of the Brillouin zone and its high-symmetry points and lines.

For model $A$, Fig.\ \ref{fig_2}(b) shows the little groups of the high-symmetry points and lines. We can now use Tables \ref{table:mono_gray} and \ref{table:di} to analyze the nodal structure in the vicinity of each of these elements. First, all high-symmetry points have to be Weyl points, shown as red dots in Fig.\ \ref{fig_2}(b). Second, none of the high-symmetry elements supports nodal lines. Third, for the points $\Gamma$ and $R$, the little group is the full group $O(T)$. According to Table \ref{table:di}, we expect four additional Weyl points in their vicinity, which lie on the four $\langle 111\rangle$ axes and are related by symmetry operations from $O(T)$, i.e., form the corners of a regular tetrahedron. These points are shown as cyan dots in Fig.\ \ref{fig_2}(b). The results for high-symmetry lines are consistent with these conclusions. For example, the line $\Lambda$ has the little group $D_3(C_3)$, for which we expect zero or two Weyl points on axis. For a general point on this line, there are no Weyl points in the vicinity. For a point close to $\Gamma$ or $R$, we find a pair: the Weyl point at $\Gamma$ or $R$ and the one out of four points on this particular axis. Moreover, for $D_3(C_3)$ we expect zero or six Weyl points off axis. More detailed analysis shows that these points form two triples with $C_3$ symmetry. The two triples are not related by any symmetry. Hence, for a particular point $\bm{k}$ along $\Lambda$, there are either zero or three Weyl points in a plane normal to $\Lambda$ and intersecting it in $\bm{k}$. The potential three points are the cyan ones that do not lie on the same axis.

All of these conclusions agree with a tight-binding Hamiltonian with point group $O(T)$ (magnetic space group \#207.3.1544). The coefficients of the Pauli matrices in Eq.~\eqref{eq:general_twoband_Hamiltonian} are given by
\begin{align}
	f_x^A(\bm{k})&= a_A \sin k_x + b_A \sin k_y \sin k_z \label{eq:O_T_x},\\
 f_y^A(\bm{k})&= a_A \sin k_y + b_A \sin k_z \sin k_x \label{eq:O_T_y},\\
	f_z^A(\bm{k})&= a_A \sin k_z + b_A \sin k_x \sin k_y \label{eq:O_T_z},
\end{align}
where $a_A, b_A\in\mathbb{R}\setminus \{0\}$. As one can check, Eq.~\eqref{eq:band_touchings_general} is satisfied at all high-symmetry points. Moreover, half of the space diagonal is parameterized by $\bm{k}_A=(t,t,t)$ with $t\in (0,\pi)$ so that Eq.~\eqref{eq:band_touchings_general} is satisfied if 
\begin{equation}\label{eq:O_T_space_diagonal}
	\sin t = -\frac{a_A}{b_A}.
\end{equation}
Equation \eqref{eq:O_T_space_diagonal} has either no solution or two solutions, which are symmetrical with respect to $t=\pi/2$ as depicted in Fig.~\ref{fig_2}(b). In the latter situation, one of these points is one of the four space-diagonal Weyl points of $\Gamma$ and the other belongs to $R$. Note that these Weyl points could annihilate; this cannot happen for the continuum model.

As a second example we consider model $B$, which belongs to point group $O_h(T_h)$ (magnetic space group \#221.4.1597). In Fig.~\ref{fig_2}(c), the little groups of the high-symmetry points and lines are depicted. Since both $\Gamma$ and $R$ have the full point-group symmetry $O_h(T_h)$ nodal lines are found on the high-symmetry axes $\Delta$ and $T$, see Table \ref{table:di}. Moreover, the little group for both $X$ and $M$ is $D_{4h}(D_{2h})$ so that there is also a nodal line on the high-symmetry axis $Z$. The presence of these nodal lines as well as the absence of nodal lines on $\Lambda$ and $\Sigma$ are consistent with the little groups of the high-symmetry line. The coefficients of the Pauli matrices for model $B$ are given by 
\begin{align}
	f_x^B(\bm{k})&= a_B \sin k_y \sin k_z \label{eq:Oh_Th_x},\\
	f_y^B(\bm{k})&= a_B \sin k_z \sin k_x \label{eq:Oh_Th_y},\\
	f_z^B(\bm{k})&= a_B \sin k_x \sin k_y \label{eq:Oh_Th_z},
\end{align}
where $a_B\in\mathbb{R}\setminus \{0\}$. It is obvious that Eq.~\eqref{eq:band_touchings_general} leads to exactly the same nodal lines as obtained from the symmetry analysis.

\section{Conclusions}

By investigating pseudospin-1/2 Hamiltonians with spin-orbit coupling in the vicinity of a reference point in momentum space, typically a high-symmetry point, we have classified twofold degenerate band touchings for the 122 magnetic point groups. Here, the pseudospin-1/2 degree of freedom is assumed to transform like a true spin-1/2 but need not actually derive from a true spin. In particular, this applies to all point groups that have only a single double-valued real irrep such as $D_2$ and $C_{2v}$ \citep{Altmann_Herzig_1994}.
How a spin-1/2 transforms under any given magnetic point group is determined by viewing the corresponding structural point group as a subgroup of $\mathrm{SU}(2)$. Bands characterized by local degrees of freedom that do not transform like a spin-1/2 can be analyzed analogously. To that end, one has to determine the real irreducible coreps of the basis matrices acting on the Hilbert space of the local degrees of freedom, in analogy to the matrices $\sigma_\mu$ in Sec.~\ref{sec:method}.

The results of our investigation are shown in Table \ref{table:mono_gray} for the gray and monochromatic groups and in Table \ref{table:di} for the dichromatic groups. We see that every magnetic point group not containing the product of time-reversal and inversion symmetries can in principle give rise to topologically nontrivial band touchings, i.e., Weyl points or nodal lines. This situation is thus common and, in particular, is realized already for the simplest possible type of multiband models, namely those characterized by a pseudospin-1/2. Noting that the $\Gamma$ point of any material has the full symmetry of the magnetic point group our results can be summarized as follows. For gray point groups, our work emphasizes the conclusion drawn by Xie \textit{et al.}\ \cite{Xie_Gao_21}: All noncentrosymmetric nonmagnetic (semi\nobreakdash-) metals are topological in this sense. Furthermore, as revealed by our analysis, any material that belongs to a monochromatic point group  is topological if the point group is either chiral and has more than one rotation axis or if it is achiral and has at least one rotation axis that is located in a mirror plane. For dichromatic point groups $\bm{G}(\bm{H})$, we find that materials that lack the product of time-reversal and inversion symmetry are topological if (i) $\bm{G}$ and $\bm{H}$ are chiral groups and have both only one rotation axis or if both have multiple rotation axes, (ii) if the product of time-reversal symmetry and an improper rotation is present, or (iii) if the halving subgroup $\bm{H}$ is either $C_{2v}$ or $C_{3v}$. 

We have seen that the existence and location of band-touching points is not influenced by terms in the Hamiltonian that are proportional to the identity matrix. However, these terms determine the energy at which the band touching occurs and are, of course, material dependent. Experiments at low energies, for example dc and low-frequency transport measurements, will only be affected by band touchings close to the Fermi energy. Finally, we wish to comment on off-axis Weyl points. Such Weyl points are missed when traversing the Brillouin zone along standard high-symmetry paths, either in experiments such as angle-resolved photoemission spectroscopy or in band-structure calculations. Our method can help to determine whether it is worthwhile to search for such points.

To conclude, we have provided a catalog that should be useful for the design and identification of topological semimetals. In particular, our work advances the active field of magnetically ordered semimetals. The method can be extended to systems of bands that are not described by a pseudospin-1/2.

\textit{Note added.}---A recent article by Liu \textit{et al.}\ \cite{Liu_Zhang_2022} catalogs emergent particles for the type-III magnetic space groups. The analysis is based on double-valued corepresentations for the Bloch states and does not predict band touchings away from high-symmetry elements.

\begin{acknowledgments}

Financial support by the Deut\-sche Forschungsgemeinschaft via the Collaborative Research Center SFB 1143, Project No.~A04, and the Cluster of Excellence on Complexity and Topology in Quantum Matter ct.qmat (EXC 2147) is gratefully acknowledged.

\end{acknowledgments}

\appendix

\section{Corepresentations}
\label{app:coreps}

In this Appendix, we first briefly review the basics of corepresentation theory, which was originally developed by Wigner \cite{Wigner_1931} and subsequently extended \cite{Dimmock_Wheeler_1962, Cracknell_Wong_1967, Cracknell_1965,Bradley_Davies_1968,Bradley_Cracknell_1972}. This theory deals with \emph{complex} coreps. It will turn out that we need to work with \emph{real} corepresentations \cite{Dyson_1962,Rumynin_Taylor_2021}. We then apply the results to determine the character tables of gray and dichromatic point groups. To clarify our notation, we will call a group element \emph{antiunitary} if it contains the time reversal $\Theta$ since the time-reversal operator is antiunitary and the product of a unitary and an antiunitary operator is again antiunitary. All group elements that solely correspond to crystalline symmetry operations are referred to as \emph{unitary}. 

Monochromatic point groups do not contain any antiunitary elements and are identical to the corresponding crystallographic point groups. Consequently, standard representation theory applies, the results of which can be found in tables and books \cite{Altmann_Herzig_1994,bilbao_server}.

\subsection{Basics of corepresentation theory}

Any gray or dichromatic magnetic point group can be written as
\begin{equation}\label{def:mag_point_grp_general}
\bm{M}=\bm{H}+A \bm{H},
\end{equation}
where $\bm{H}$ is the subgroup of $\bm{M}$ that only contains the unitary elements. In Eq.~\eqref{def:mag_point_grp_general}, $A$ denotes an arbitrary but fixed antiunitary element of $\bm{M}$. Gray groups contain $\Theta$ and so the choice $A=\Theta$ is possible and common but not mandatory. In the following, we denote unitary elements of $\bm{H}$ by $R$ and $S$, and antiunitary elements of the left coset $A \bm{H}$ by $B$ and $C$, respectively. Since $\bm{H}$ is a group the action of a unitary element $R$ on a basis function $\psi^\nu = \left(\psi_1^\nu,\psi_2^\nu,\dots,\psi_n^\nu\right)$ of the $n$-dimensional representation $\nu$ is given by
\begin{equation}\label{def:basis_fct_unitary_elements}
\hat{P}_R\psi^\nu  = \psi^\nu \Delta^\nu(R),
\end{equation}
with the matrix representative $\Delta^\nu(R)$ of $R$. By acting with $A$ on the basis function $\psi^\nu$, we obtain a new set of basis functions 
\begin{equation}
\phi^\nu = \hat{P}_A\psi^\nu, 
\end{equation}
with $\phi^\nu=(\phi_1^\nu,\phi_2^\nu,\dots,\phi_n^\nu)$. The functions $\psi^\nu$ and $\phi^\nu$ are not necessarily  linearly dependent since $A$ is antiunitary. Hence, a more suitable basis function for the representation of $\bm{M}$ is given by 
\begin{equation}
\gamma^\nu = \left(\psi^\nu,\phi^\nu\right)=\left(\psi_1^\nu,\psi_2^\nu,\dots,\psi_n^\nu,\phi_1^\nu,\phi_2^\nu,\dots,\phi_n^\nu\right). 
\end{equation} 
For unitary elements $R\in\bm{H}$, the matrix representation corresponding to $\gamma^\nu$ is written as
\begin{equation}\label{eq:def_corep_uni}
D^\nu(R)=\left(\begin{matrix}
\Delta^\nu(R) & 0 \\
0 & \Delta^\nu(A^{-1} R A)^*
\end{matrix}\right),
\end{equation}
whereas for antiunitary elements $B\in A\bm{H}$, we have
\begin{equation}\label{eq:def_corep_antiuni}
D^\nu(B)=\left(\begin{matrix}	\hiderowcolors
0 & \Delta^\nu(BA) \\
\Delta^\nu(A^{-1} B)^* & 0
\end{matrix}\right).
\end{equation}
The set of unitary matrices defined by Eqs.\ \eqref{eq:def_corep_uni} and \eqref{eq:def_corep_antiuni} are called the \emph{complex coreps} of $\bm{M}$ derived from $\Delta^\nu$. In contrast to ordinary representations that are ho\-mo\-mor\-phisms, coreps are not struc\-ture-pre\-ser\-ving maps, as reflected by the product rules 
\begin{align}
D(R S) &= D(R) D(S),\label{eq:calc_rules_coreps_1}\\
D(R B) &= D(R) D(B),\label{eq:calc_rules_coreps_2}\\
D(B S) &= D(B) D(S)^*,\label{eq:calc_rules_coreps_3}\\
D(B C) &= D(B) D(C)^*.\label{eq:calc_rules_coreps_4}
\end{align}
Note the complex conjugation of the matrix representatives $D(S)$ and $D(C)$ in Eqs.~\eqref{eq:calc_rules_coreps_3} and \eqref{eq:calc_rules_coreps_4}, respectively,  which follow the matrix representative $D(B)$ of the antiunitary element $B$.

Next, we investigate how coreps behave with respect to unitary transformations. By performing a unitary transformation $U$, the basis function is transformed according to
\begin{equation}
\gamma'^\nu=\gamma^\nu U, 
\end{equation} 
whereas for the coreps, we have
\begin{align}
D'^\nu(R) &= U^{\dagger} D^\nu(R)\, U,\label{eq:coreps_uni_trafo_uni}\\
D'^\nu(B) &= U^{\dagger} D^\nu(B)\, U^*.\label{eq:coreps_uni_trafo_antiuni}
\end{align}
In the standard theory, the notion of reducibility of coreps is defined like for ordinary representations: If we can find a unitary transformation that brings all coreps $D^\nu(R)$ and $D^\nu(B)$ of $\bm{M}$ into block-diagonal form, then the corep is called reducible. In other words, a corep $D^\nu$ is reducible if it is unitarily equivalent to a block-diagonal form. Conversely, if no such unitary transformation  exists the corep is said to be irreducible. 

For our purposes, the concept of unitary equivalence and the corresponding definition of reducibility is not appropriate. Rather, we need to restrict ourselves to real unitary, i.e., orthogonal, transformations for the following reason. The antiunitary elements of the gray and dichromatic point groups are antilinear, i.e., they act on complex numbers as complex conjugation. As an illustration of such an antiunitary element, we consider the time-reversal operator $\mathcal{T}$, which can be written as
\begin{equation}
\mathcal{T}=U_\mathcal{T} K,
\end{equation} 
where $U_\mathcal{T}$ is  unitary and $K$ is the operator of complex conjugation defined with respect to a certain basis of the Hilbert space. $K$ leaves elements of this basis invariant but replaces all numbers by their complex conjugates. Now, if we perform a general unitary transformation on the corep we mix the real and imaginary parts of the components of the matrix representatives of the corep. This would require us to also change the operator $K$ of complex conjugation. Another consequence is that we would not be able to distinguish coreps that are even or odd under time reversal. This is seen by transforming $D^\nu(\Theta)$ with $U = i \bbone$, where $\bbone$ is the identity matrix,
\begin{equation}
	D^{\prime\nu}(\Theta) = U^\dagger D^\nu(\Theta) U^* = -i D^\nu(\Theta)(-i) = - D^\nu(\Theta) .
\end{equation}
Here, coreps $D^\nu$ and $D^{\prime\nu}$ have opposite parity under time reversal but are unitarily equivalent. Since we need to distinguish between time-reversal-even and time-reversal-odd coreps unitary equivalence is not the appropriate notion of equivalence. This problem is avoided by instead employing only real orthogonal transformations. Accordingly, we call two (real) coreps \emph{real equivalent} if there is an orthogonal transformation connecting them and we call a corep \emph{real irreducible} if there is no orthogonal transformation that block-diagonalizes all matrix representatives.
In Appendices \ref{app.corep.gray} and \ref{app.corep.di}, we will show how to derive the character tables for single-valued real coreps of gray and dichromatic point groups, and we will further demonstrate the necessity to use orthogonal transformations instead of unitary ones.

\subsection{Corepresentation theory for single-valued coreps of gray point groups}
\label{app.corep.gray}

In this Appendix, we show how to calculate the character table for single-valued real coreps of gray point groups. To simplify the notation, we will suppress the superscript $\nu$ that specifies the representation. For gray point groups, it is convenient to chose $A=\Theta$. Then, the coreps of the unitary elements are given by
\begin{align}
D(R)&= \left(\begin{matrix}	
\Delta(R) & 0 \\
0 & \Delta(A^{-1} R A)^*
\end{matrix}\right) = \left(\begin{matrix}
\Delta(R) & 0 \\
0 & \Delta(\Theta^{-1} R \Theta)^*
\end{matrix}\right) \nonumber \\
&=  \left(\begin{matrix}	
\Delta(R) & 0 \\
0 & \Delta(\Theta^{-1} \Theta R )^*
\end{matrix}\right)=\left(\begin{matrix}
\Delta(R) & 0 \\
0 & \Delta(R)^*
\end{matrix}\right)\label{eq:corep_gray_uni}. 
\end{align}    
The antiunitary elements $B$ of a gray point group are elements of the left coset $\Theta \bm{H}$ and can be written as $B=\Theta R'$, where $R' \in \bm{H}$. Hence, the coreps of antiunitary elements are given by 
\begin{align}
D(B) &= D(\Theta R') = \left(\begin{matrix}	
0 & \Delta(\Theta R' \Theta) \\
\Delta(\Theta^{-1} \Theta R')^* & 0
\end{matrix}\right) \nonumber \\
&= \left(\begin{matrix}	
0 & \Delta(\Theta^2 R') \\
\Delta(R')^* & 0
\end{matrix}\right)
= \left(\begin{matrix}	
0 & \Delta(\tilde{e} R') \\
\Delta(R')^* & 0
\end{matrix}\right) \nonumber \\
&= \left(\begin{matrix}	
0 & \Delta(R') \\
\Delta(R')^* & 0
\end{matrix}\right).\label{eq:corep_gray_antiuni}
\end{align}
In the derivation, we have used that $\Theta^2 = \tilde{e}$ \cite{Wigner_1931}, where $\tilde{e}$ is the group element that corresponds to rotations of $2\pi$. It satisfies $\tilde{e}^2 = e$ and is mostly relevant for double-valued (spinor) coreps. Here, we are only interested in single-valued coreps so that $\Delta(\tilde{e})=\Delta(e)$ holds. Thus, we identify $e$ and $\tilde{e}$ with each other since only the matrix representatives are relevant for our purposes. Using Eq.~\eqref{eq:corep_gray_antiuni}, the corep of $\Theta$ for an $n$-dimensional single-valued representation is obtained as 
\begin{align}
D(\Theta) &= D(\Theta e)
= \left(\begin{matrix}	
0 & \Delta(e)\\
\Delta(e)^* & 0\\ \end{matrix}\right)
= \left(\begin{matrix}	
0 & \bbone_n\\
\bbone_n & 0\\ \end{matrix}\right) \nonumber \\
&= \left(\begin{matrix}0 & 1 \\
1 & 0
\end{matrix}\right)\otimes \bbone_n,\label{eq:corep_time_rev}
\end{align} 
where $\bbone_n$ is the $n$-dimensional identity matrix. The matrix $D(\Theta)$ is real and symmetric, and thus orthogonally diagonalizable. For $n=1$, we have $\bbone_1=1$ and the orthogonal transformation that diagonalizes $D(\Theta)$ is given by 
\begin{equation}\label{eq:uni_trafo_U1}
U^1 =\frac{1}{\sqrt{2}}\left( \begin{matrix}
-1 & 1 \\
1 & 1 \\
\end{matrix}\right).  
\end{equation}
Thus, for arbitrary dimension $n$, the orthogonal transformation matrix $U^n_g$ that diagonalizes $D(\Theta)$ is
\begin{equation}\label{eq:Un}
U^n_g = U^1 \otimes \bbone_n . 
\end{equation}
By applying $U^n_g$ to $D(\Theta)$, we find 
\begin{equation}\label{eq:corep_theta_diag}
D'(\Theta)=\left(U^n_g\right)^{\dagger} D(\Theta)\, U^n_g= \left(\begin{matrix}	
-\bbone_n & 0 \\
0 & \bbone_n
\end{matrix}\right), 
\end{equation}
where the corep in which the matrix representative of $\Theta$ is diagonal is denoted by  a prime. Next, we apply $U^n_g$ to the coreps of the unitary elements $D(R)$. In this work, the subgroup $\bm{H}$ of unitary elements is identical to one of the 32 crystallographic point groups. Then, for single-valued representations, all matrix representations of the unitary elements $\Delta(R)$ can be assumed to be real if we treat one-dimensional complex-irreducible representations, which occur in pairs of complex conjugates, such as $^1E$ and $^2E$ for the point group $C_3$, as two-dimensional real-irreducible representations. For real representations, we have $\Delta(R)^*=\Delta(R)$ and Eq.~\eqref{eq:corep_gray_uni} can be written as 
\begin{equation}
D(R)=\bbone_2\otimes \Delta(R). 
\end{equation}   
Next, we apply $U^n_g$ to $D(R)$ and find
\begin{align}
D'(R) = \left(U^n_g\right)^{{\dagger}} D(R)\, U^n_g = \left(\begin{matrix}	
\Delta(R) & 0 \\
0 & \Delta(R)
\end{matrix}\right). \label{eq:gray_D_R_diag}
\end{align}
Obviously, $D'(R)$ is block diagonal for all $R\in\bm{H}$. In order to calculate the coreps of the antiunitary elements $B\in A\bm{H}$, we employ Eq.~\eqref{eq:calc_rules_coreps_3} and use that we can express every antiunitary element as $B=\Theta R'$. The result is
\begin{equation}
D'(B)=D'(\Theta R') =D'(\Theta) D'(R')=\left(\begin{matrix}	\hiderowcolors
-\Delta(R') & 0 \\
0 & \Delta(R')
\end{matrix}\right),\label{eq:gray_D_B_diag}
\end{equation}  
where we have further used that $D'(R')$ is real.
From Eqs.~\eqref{eq:gray_D_R_diag} and \eqref{eq:gray_D_B_diag}, we find that the coreps of gray point groups are always reducible to the two real-irreducible coreps, where one corep is odd and the other is even under time reversal, as reflected by Eq.~\eqref{eq:corep_theta_diag}.

\begin{table}
	\begin{center}
		\caption{Character table of the monochromatic point group $D_2$ with basis functions up to second order in momentum and spin basis functions.}
		\begin{tabular}{c c  c  c  c  c }
			\hline \hline
			$D_2$ & $e$ & $C_{2z}$ & $C_{2x}$ & $C_{2y}$ & basis functions \\ \hline
			\rule{0pt}{3ex}
			$A$ & $1$ & $1$ & $1$ & $1$ & $1$, $k_x^2$, $k_y^2$, $k_z^2$ \\
			\rule{0pt}{3ex}
			$B_1$ & $1$ & $1$ & $-1$ & $-1$ & $\sigma_z$, $k_z$, $k_x k_y$ \\
			\rule{0pt}{3ex}
			$B_2$ & $1$ & $-1$ & $-1$ & $1$ & $\sigma_y$, $k_y$, $k_z k_x$ \\
			\rule{0pt}{3ex}
			$B_3$ & $1$ & $-1$ & $1$ & $-1$ & $\sigma_x$, $k_x$, $k_y k_z$ \\
			\hline \hline
		\end{tabular}
		\label{table:char_tab_D2}
	\end{center} 	
\end{table}  

It remains to calculate basis functions of $D'$ based on the basis functions of $\Delta$. For single-valued representations, products of powers of components of the momentum vector can be chosen as basis functions. We here also treat the identity matrix $\sigma_0$ and the Pauli matrices $\sigma_i$ as ``spin basis functions,'' as it is often done \cite{Pauli_1927, Dresselhaus_2007, Chiu_Teo_2016}. They are more correctly irreducible tensor operators. First, we consider those basis functions of the $n$-dimensional representation $\Delta$ that are odd under time reversal, such as products of odd order of momentum components and the Pauli matrices, and denote them by $\psi^-$. Then, the basis function of the corep $D$ is given by
\begin{equation}
\gamma^- = (\psi^- ,\hat{P}_A \psi^-)=(\psi^- ,\hat{P}_\Theta \psi^-)=(\psi^- ,- \psi^-).  
\end{equation} 
In order to find the basis function of $D'$, we perform the orthogonal transformation $U^n_g$,
\begin{equation}
\gamma'^- = \gamma^- U^n_g = \frac{1}{\sqrt{2}}(- \psi^- - \psi^-, \psi^- - \psi^-)
= (-\sqrt{2}\, \psi^-, 0). 
\end{equation}
Thus, $\psi^-$ is a basis function of the real-irreducible corep that is odd under time reversal. Analogously, we find that those basis functions of $\Delta$ that are even under time-reversal, i.e., products of even order of momentum components, belong to the real-irreducible corep that is even under time reversal. Hence, all basis functions behave under time reversal as expected.

\begin{table*}
	\begin{center}
		\caption{Character table of the gray point group $D_2\otimes\{e,\Theta\}$ with basis functions up to second order in momentum and spin basis functions. The lowest-order basis function of $A^-$ is $k_xk_yk_z$.}
		
		\begin{tabular}{c c  c  c  c  c  c  c  c c }
			\hline \hline
			$D_2\otimes\{e,\Theta\}$ & $e$& $C_{2z}$ & $C_{2x}$ & $C_{2y}$ & $\Theta$ & $\Theta C_{2z}$ & $\Theta C_{2x}$ & $\Theta C_{2y}$ & basis functions\\\hline
			\rule{0pt}{3ex}
			$A^+$ & $1$ & $1$ & $1$ & $1$ & $1$ & $1$ & $1$ & $1$ & $\sigma_0$, $1$, $k_x^2$, $k_y^2$, $k_z^2$ \\
			\rule{0pt}{3ex}
			$A^-$ & $1$ & $1$ & $1$ & $1$ & $-1$ & $-1$ & $-1$ & $-1$ & \\
			\rule{0pt}{3ex}
			$B_1^+$ & $1$ & $1$ & $-1$ & $-1$ & $1$ & $1$ & $-1$ & $-1$ & $k_x k_y$ \\
			\rule{0pt}{3ex}
			$B_1^-$ & $1$ & $1$ & $-1$ & $-1$ & $-1$ & $-1$ & $1$ & $1$ & $\sigma_z$, $k_z$ \\
			\rule{0pt}{3ex}
			$B_2^+$ & $1$ & $-1$ & $-1$ & $1$ & $1$ & $-1$ & $-1$ & $1$ & $k_z k_x$ \\
			\rule{0pt}{3ex}
			$B_2^-$ & $1$ & $-1$ & $-1$ & $1$ & $-1$ & $1$ & $1$ & $-1$ & $\sigma_y$, $k_y$ \\
			\rule{0pt}{3ex}
			$B_3^+$ & $1$ & $-1$ & $1$ & $-1$ & $1$ & $-1$ & $1$ & $-1$ & $k_y k_z$ \\
			\rule{0pt}{3ex}
			$B_3^-$ & $1$ & $-1$ & $1$ & $-1$ & $-1$ & $1$ & $-1$ & $1$ & $\sigma_x$, $k_x$ \\
			\hline \hline
		\end{tabular}
		\label{table:char_tab_D2_gray}
	\end{center} 	
\end{table*} 

Next, we illustrate our results by the gray point group $D_2\otimes\{e,\Theta\}$. The character table of the point groups $D_2$ and $D_2\otimes\{e,\Theta\}$ are given in Tables \ref{table:char_tab_D2} and \ref{table:char_tab_D2_gray}, respectively. Since the order of $D_2\otimes\{e,\Theta\}$ is twice the order of $D_2$ and both only have one-dimensional real-irreducible (co-) representations there are twice as many real coreps for $D_2\otimes\{e,\Theta\}$ as there are for $D_2$. In fact, we find that each irrep of $D_2$ is split into one time-reversal-even and one time-reversal-odd real corep for $D_2\otimes\{e,\Theta\}$, denoted by the superscripts $+$ and $-$, respectively.

Let us consider what happens if we allow not only orthogonal but general unitary transformations. Then, any two coreps of $D_2\otimes\{e,\Theta\}$ that differ only in their parity with respect to $\Theta$ are equivalent. For example, let us apply the unitary transformation $U=i$ to the representation $B_1^-$. Equations \eqref{eq:coreps_uni_trafo_uni} and \eqref{eq:coreps_uni_trafo_antiuni} show that the coreps of all unitary elements stay the same whereas the ones of all antiunitary elements change sign. Hence, $B_1^-$ is unitarily equivalent to $B_1^+$. Analogous results are found for the other coreps. Thus $D_2\otimes\{e,\Theta\}$ only has the four \emph{complex} irreducible coreps $A$, $B_1$, $B_2$, and $B_3$ in accordance with Refs.~\cite{Xu_Elcoro_2020,Elcoro_Wieder_2021}. These complex coreps do not allow to distinguish between time-reversal-even and time-reversal-odd basis functions.

\subsection{Corepresentation theory for single-valued coreps of dichromatic point groups}
\label{app.corep.di}

For single-valued real coreps of dichromatic point groups, the discussion is more complicated than for gray point groups. The reason for this is that dichromatic point groups do not share a common antiunitary element that could be used in order to construct the real coreps, in contrast to gray point groups. For the majority of dichromatic point groups, one can choose the unitary part $R'$ of the antiunitary element $A=\Theta R'$ in such a way that it corresponds to a twofold symmetry, i.e., $R'^2=e$. However, for the dichromatic point groups $C_4(C_2)$, $S_4(C_2)$, and $C_{4h}(C_{2h})$ it is not possible to choose $R'$ in this way. In the following, we show how the real coreps from one-dimensional, two-di\-men\-sional, and three-dimensional representations are derived if one can choose $A=\Theta R'$ with $R'^2=e$. Furthermore, the real coreps of $S_4(C_2)$ are derived as an example for a dichromatic point group for which no antiunitary element $A$ with $R'^2=e$ exists.
	
For the first case, we choose the antiunitary element $A$ as $A=\Theta R'$ with $R'^2=e$. The coreps of the unitary elements are then calculated as 
\begin{align}
D(R) &= \left(\begin{matrix}
\Delta(R) & 0 \\
0 & \Delta(A^{-1} R A)^*
\end{matrix}\right) \nonumber \\
&= \left(\begin{matrix}	
\Delta(R) & 0 \\
0 & \Delta((\Theta R')^{-1} R (\Theta R'))^*
\end{matrix}\right) \nonumber \\
&= \left(\begin{matrix}
\Delta(R) & 0 \\
0 & \Delta(R'^{-1} R R')^*
\end{matrix}\right),\label{eq:corep_di_unitary}
\end{align}
where we have used that $\Theta$ commutes with all group elements \cite{Wigner_1931,Bradley_Cracknell_1972}. Since $R'^{-1} R R'$ is an element of the conjugacy class of $R$ the matrix representatives $\Delta(R)$ and $\Delta(R'^{-1}R R')$ have the same trace, which means that they are identical for one-dimensional representations. For the corep of the antiunitary element $A=\Theta R'$, we obtain
\begin{align}
D(A) &= \left(\begin{matrix}
0 & \Delta(AA) \\
\Delta(A^{-1} A)^* & 0
\end{matrix}\right)
= \left(\begin{matrix}
0 & \Delta(\Theta R' \Theta R') \\
\Delta(e)^* & 0
\end{matrix}\right) \nonumber \\
&= \left(\begin{matrix}
0 & \Delta(\Theta^2 R'^2) \\
\Delta(e)^* & 0
\end{matrix}\right)=\left(\begin{matrix}
0 & \Delta(\tilde{e} e) \\
\Delta(e)^* & 0
\end{matrix}\right)\nonumber\\
&=\left(\begin{matrix}
0 & \Delta(e) \\
\Delta(e)^* & 0
\end{matrix}\right)=\left(\begin{matrix}
0 & \bbone_n \\
\bbone_n & 0
\end{matrix}\right).\label{eq:corep_di_antiunitary}
\end{align} 
Next, let us consider the conjugacy classes of the monochromatic point group $\bm{G}$ and the dichromatic point group $\bm{G}(\bm{H})$, where we regard $\bm{G}$ and $\bm{G}(\bm{H})$ as subgroups of the gray group $\bm{G}\otimes\{e,\Theta\}$. Moreover, $\bm{H}$ is a halving subgroup of $\bm{G}$. Let $R_1\in\bm{H}$ and $\tilde{R}_1\in \bm{G}-\bm{H}$ be fixed elements. Then, the conjugacy class of $R_1$ is given by 
\begin{align}
K(R_1) &= \{R_2^{-1}R_1 R_2\,|\, R_2\in\bm{H}\} \nonumber \\
&\quad{} \cup \{\tilde{R}_2^{-1}R_1 \tilde{R}_2\,|\, \tilde{R}_2\in\bm{G}-\bm{H}\}
\label{eq:conj_class_1}
\end{align}  
and the conjugacy class of $\tilde{R}_1$ reads as
\begin{align}
K(\tilde{R}_1) &= \{R_2^{-1}\tilde{R}_1 R_2\,|\, R_2\in\bm{H}\} \nonumber \\
&\quad{} \cup \{\tilde{R}_2^{-1}\tilde{R}_1 \tilde{R}_2\,|\, \tilde{R}_2\in\bm{G}-\bm{H}\}.
\label{eq:conj_class_2} 
\end{align}
For the conjugacy class of the unitary element $R_1 \in \bm{H} \subset \bm{G}(\bm{H})$, we use that $(\Theta\tilde{R}_2)^{-1}R_1 \Theta\tilde{R}_2=\tilde{R}_2^{-1}R_1 \tilde{R}_2$ due to the commutativity of $\Theta$ with all other group elements. Thus, $K(R_1)$ is a conjugacy class of $\bm{G}$ and $\bm{G}(\bm{H})$. The conjugacy class of $\Theta\tilde{R}_1\in\Theta(\bm{G}-\bm{H})$ is calculated as
\begin{align}
K(\Theta\tilde{R}_1) &= \{R_2^{-1}\Theta\tilde{R}_1 R_2\,|\, R_2\in\bm{H}\} \nonumber \\
&\quad{} \cup \{(\Theta\tilde{R}_2)^{-1}\Theta\tilde{R}_1 \Theta\tilde{R}_2\,|\, \Theta\tilde{R}_2\in\Theta(\bm{G}-\bm{H})\}\nonumber\\
&= \{R_2^{-1}\Theta\tilde{R}_1 R_2\,|\, R_2\in\bm{H}\} \nonumber \\
&\quad{} \cup \{\tilde{R}_2^{-1}\Theta\tilde{R}_1 \tilde{R}_2\,|\,\tilde{R}_2\in\bm{G}-\bm{H}\} \nonumber\\
&= \Theta\big(\{R_2^{-1}\tilde{R}_1 R_2\,|\, R_2\in\bm{H}\} \nonumber \\
&\quad{} \cup \{\tilde{R}_2^{-1}\tilde{R}_1 \tilde{R}_2\,|\,\tilde{R}_2\in\bm{G}-\bm{H}\}\big) \nonumber \\
&= \Theta K(\tilde{R}_1).\label{eq:conj_class_3} 
\end{align}
Hence, if $K(\tilde{R}_1)$ is a conjugacy class of $\bm{G}$, then $\Theta K(\tilde{R}_1)$ is a conjugacy class of $\bm{G}(\bm{H})$ as shown in Eq.~\eqref{eq:conj_class_3}. Consequently, $\bm{G}$ and $\bm{G}(\bm{H})$ have the same number of conjugacy classes. In the following sections, we apply the just obtained results to one-, two-, and three-dimensional representations to derive the real coreps.

\subsubsection{Coreps of dichromatic point groups derived from real one-dimensional representations}

In this section, we consider a corep $D^1$ that is derived from a real one-dimensional representation $\Delta^1$, i.e., we consider only $A$ or $B$ representations but not complex representations such as $^1E$ and $^2E$ of the monochromatic point group $C_4$. We will include one-dimensional complex representations later in the discussion of two-dimensional real representations. Since we are only taking into account real one-dimensional representations, we have $\Delta^1(R)=\Delta^1(R'^{-1}RR')=\pm 1$, where the sign depends on the specific  $R\in\bm{H}$. Thus, with Eq.~\eqref{eq:corep_di_unitary} the corep for a unitary element $R$ is given by
\begin{equation}
	D^1(R)=\pm\left(\begin{matrix}
	1 & 0 \\
	0 & 1
	\end{matrix}\right).
\end{equation}
For $A$, the corep follows from Eq.~\eqref{eq:corep_di_antiunitary},
\begin{equation}
	D^1(A)=\left(\begin{matrix}
	0 & 1 \\
	1 & 0
	\end{matrix}\right). 
\end{equation}
Thus, $D^1$ is block diagonalized by $U^1$ in Eq.~\eqref{eq:uni_trafo_U1}. Using Eqs.~\eqref{eq:coreps_uni_trafo_uni} and \eqref{eq:coreps_uni_trafo_antiuni}, the corep $D'^1$ for which all matrices are block-diagonal is found as
\begin{equation}\label{eq:corep_R_di_1D}
D'^1(R)=\pm\left(\begin{matrix}
1 & 0 \\
0 & 1
\end{matrix}\right)
\end{equation}
and
\begin{equation}\label{eq:corep_A_di_1D}
D'^1(A)=\left(\begin{matrix}
-1 & 0 \\
0 & 1
\end{matrix}\right). 
\end{equation}
The remaining coreps of the antiunitary elements $B\in A\bm{H}$ are obtained using Eq.~\eqref{eq:calc_rules_coreps_3}. By inspection of Eqs.~\eqref{eq:corep_R_di_1D} and \eqref{eq:corep_A_di_1D}, we find that $D'^1$ is reducible into two real coreps
$d^1_-$ and $d^1_+$,
\begin{equation}\label{eq:corep_D1_direct_sum}
	D'^1 = d^1_- \oplus d^1_+, 
\end{equation}
where $d^1_-$ and $d^1_+$ are odd and even with respect to $A$, respectively. Both coreps have in common that they share the same characters for the unitary elements $R\in\bm{H}$ but differ in the sign of the characters of the antiunitary elements. 

In order to construct the basis functions of $D^1$, we differentiate between basis functions of $\Delta^1$ that are odd and even with respect to $\Theta$. They are denoted by $\psi^o_1$ and $\psi^e_1$, respectively. To simplify the following discussion, we assume that $\Delta^1(R')=1$. The case $\Delta^1(R')=-1$ can be treated analogously. For $\Delta^1(R')=1$, the basis functions of $D^1$ are given by
\begin{align}
\gamma^o_1 &= (\psi^o_1 ,\hat{P}_A \psi^o_1)=(\psi^o_1 ,\hat{P}_{\Theta R'} \psi^o_1)=(\psi^o_1 ,\hat{P}_{\Theta}\hat{P}_{R'} \psi^o_1)\nonumber\\
&=(\psi^o_1 ,\hat{P}_{\Theta}(+1)\psi^o_1)=(\psi^o_1 ,-\psi^o_1)
\end{align}
and 
\begin{align}
\gamma^e_1 &= (\psi^e_1 ,\hat{P}_A \psi^e_1)=(\psi^e_1 ,\hat{P}_{\Theta R'} \psi^e_1)=(\psi^e_1 ,\hat{P}_{\Theta}\hat{P}_{R'} \psi^e_1)\nonumber\\
&=(\psi^e_1 ,\hat{P}_{\Theta}(+1)\psi^e_1)=(\psi^e_1 ,\psi^e_1).
\end{align}
Hence, the basis functions of $D'^1$ are
\begin{equation}
\gamma'^o_1 = \gamma^o_1 U^1 = \frac{1}{\sqrt{2}}(- \psi^o_1 - \psi^o_1, \psi^o_1 - \psi^o_1)
= -\sqrt{2}\, (\psi^o_1, 0) \label{eq:basis_fct_di_1}
\end{equation}
and
\begin{equation}
\gamma'^e_1 = \gamma^e_1 U^1 = \frac{1}{\sqrt{2}}(- \psi^e_1 + \psi^e_1, \psi^e_1 + \psi^e_1)
= \sqrt{2}\, (0, \psi^e_1). \label{eq:basis_fct_di_2}
\end{equation}
By comparison of Eqs.~\eqref{eq:corep_R_di_1D}, \eqref{eq:corep_A_di_1D}, \eqref{eq:basis_fct_di_1}, and \eqref{eq:basis_fct_di_2}, we find that $\psi_1^e$ and $\psi^o_1$ are basis functions of  $d^1_+$ and $d^1_-$, respectively. In particular, $\psi_1^e$, which is an even basis function with respect to $\Theta$ and belongs to the irrep with $\Delta^1(R')=1$, is also a basis function of the real corep with $d_+^1(\Theta R')=1$. 

\begin{table}
	\begin{center}
		\caption{Character table of the dichromatic point group $D_2(C_2)$ with basis functions up to second order in momentum and spin basis functions.}
		\begin{tabular}{c c  c  c  c  c }
			\hline \hline
			$D_2(C_2)$ & $e$ & $C_{2z}$ & $\Theta C_{2x}$ & $\Theta C_{2y}$ & basis functions\\\hline
			\rule{0pt}{3ex}
			$A$ & $1$ & $1$ & $1$ & $1$ & $1$, $\sigma_z$, $k_z$, $k_x^2$, $k_y^2$, $k_z^2$ \\
			\rule{0pt}{3ex}
			$B_1$ & $1$ & $1$ & $-1$ & $-1$ & $k_x k_y$ \\
			\rule{0pt}{3ex}
			$B_2$ & $1$ & $-1$ & $-1$ & $1$ & $\sigma_x$, $k_x$, $k_z k_x$ \\
			\rule{0pt}{3ex}
			$B_3$ & $1$ & $-1$ & $1$ & $-1$ & $\sigma_y$, $k_y$, $k_y k_z$ \\
			\hline \hline
		\end{tabular}
		\label{table:char_tab_D2_C2}
	\end{center} 	
\end{table}  

Motivated by this and the discussion of the conjugacy classes in the previous section, we introduce the labeling for the real one-dimensional coreps as follows. Since the halving subgroup $\bm{H}$ of $\bm{G}$ is a monochromatic point group itself, the one-dimensional representations of $\bm{G}$ come in pairs of two. Two one-dimensional representations of $\bm{G}$ form a pair if the characters of the conjugacy classes containing the elements of $\bm{H}$ are identical, and the characters of the conjugacy classes of the group elements in the set $\bm{G}-\bm{H}$ differ in their sign. This is similar to what we have found for $d^1_-$ and $d^1_+$ with respect to the unitary elements $R\in\bm{H}$ and the antiunitary elements $ B\in\Theta (\bm{G}-\bm{H})$. By identifying the conjugacy classes $K(R)$ with $R\in\bm{H}$ of $\bm{G}$ and $\bm{G}(\bm{H})$ and the conjugacy classes $K(\tilde{R})$ with $\tilde{R}\in\bm{G}-\bm{H}$ with the conjugacy classes of the antiunitary elements $K(\Theta\tilde{R})$ with $\Theta\tilde{R}\in\Theta(\bm{G}-\bm{H})$, we identify the irrep $\Delta^1$ of $\bm{G}$ with that real corep $d_x^1$ ($x=\pm$) of $\bm{G}(\bm{H})$, which satisfies
\begin{equation}\label{eq:identification_rep_corep_1D_di}
	\Delta^1(R')=d_x^1(\Theta R') .
\end{equation}
Note that since the characters of the elements in $\bm{H}$ are identical for $\bm{G}$ and $\bm{G}(\bm{H})$, Eq.~\eqref{eq:identification_rep_corep_1D_di} implies that the remaining characters for the elements in $\bm{G}-\bm{H}$ and $\Theta (\bm{G}-\bm{H})$ are identical as well. Thus, we can rewrite Eq.~\eqref{eq:corep_D1_direct_sum} as
\begin{equation}
	D'^1=\Delta'^1\oplus\Delta^1,
\end{equation}
where $\Delta'^1$ and $\Delta^1$ are now  understood as real coreps with $\Delta^1(\Theta R')=1$ and $\Delta'^1(\Theta R')=-1$. As an example, the character tables of $D_2$ and $D_2(C_2)$ are given in Table \ref{table:char_tab_D2} and Table \ref{table:char_tab_D2_C2}, respectively.

\subsubsection{Coreps of dichromatic point groups derived from two-dimensional representations}

For coreps derived from two-dimensional representations, the discussion is somewhat different compared to the one-dimensional case. The reason for this is that the matrix representatives of $\Delta(R)$ and $\Delta(R'^{-1}RR')$, which appear in the calculation of $D(R)$, are only identical for one-dimensional representations but not necessarily for two-dimensional ones. Note that we consider one-dimensional complex irreducible representation, which come in complex-conjugate pairs, as real two-dimensional representations. 

For a two-dimensional representation $\Delta^2$ of a monochromatic point group $\bm{G}$, the real matrix representatives can take the following two forms \cite{Bradley_Cracknell_1972}, up to an overall sign,
\begin{align}
	m^2_1 =\left(\begin{matrix}
	a & b \\
	-b & a
	\end{matrix}\right),\quad m^2_2 =\left(\begin{matrix}
	a & b \\
	b & -a
	\end{matrix}\right), \label{eq:matrix_rep_two_dim}
\end{align}
with $a=\pm{1}/{2}$ and $b=\pm{\sqrt{3}}/{2}$, $a=\pm 1$ and $b=0$, or $a=0$ and $b=\pm 1$. If a matrix representative in Eq.~\eqref{eq:matrix_rep_two_dim} squares to the identity matrix it corresponds to a twofold symmetry $R'$. Furthermore, the trace of a matrix representative of a twofold symmetry can either be $\chi(\Delta^2(R'))=0$ or $\chi(\Delta^2(R'))=\pm 2$.

First, we consider the case for which $\chi(\Delta^2(R'))=0$, i.e., $\Delta^2(R')$ is of the form $m^2_2$. In this case, we choose the two-dimensional representation from which we derive the corep in such a way that the matrix representative of the unitary part of $A=\Theta R'$ is given by 
\begin{equation}\label{eq:mat_rep_chi_0}
	\Delta^2(R')=\left(\begin{matrix}
	1 & 0 \\
	0 & -1
	\end{matrix}\right), 
\end{equation}
which is $m^2_2$ for $a=1$ and $b=0$. Note that we can always choose Eq.~\eqref{eq:mat_rep_chi_0} as the matrix representative of $R'$ if $\chi(\Delta^2(R'))=0$ since two matrices are similar if they share the same eigenvalues and the eigenvectors of each matrix are linearly independent. In order to calculate the lower block matrix of the coreps of the unitary elements $R\in\bm{H}$ in Eq.~\eqref{eq:corep_di_unitary}, we use that all matrix representatives are real and that
\begin{equation}\label{eq:corep_two_dim_unitary}
	\Delta^2(R'^{-1}RR')=\Delta^2(R')^{-1}\Delta^2(R)\Delta^2(R'). 
\end{equation}
Since $e$ is always an element of a halving subgroup $\bm{H}$ there is always at least one matrix representative, namely $\Delta^2(e)$, of type $m_1^2$. Furthermore, $\bm{H}$ can contain elements the matrix representatives of which are of type $m_2^2$. Hence, using Eq.~\eqref{eq:corep_two_dim_unitary} the coreps of the unitary elements are of the form
\begin{equation}\label{eq:corep_di_2D}
D^2_{m^2_i}(R)=\left(\begin{matrix}
m^2_i & 0 \\
0 & \Delta^2(R')^{-1}m^2_i\Delta^2(R')
\end{matrix}\right),
\end{equation} 
with $i=1,2$, and the specific form of $m^2_i$ depends on $R$. Note that both block matrices in Eq.~\eqref{eq:corep_di_2D} have the same trace $\chi(m_i^2)$. To be more precise, the corep for $m^2_1$ and $m^2_2$ are given by
\begin{align}
	D^2_{m^2_1}(R) &= \left(\begin{matrix} 
	a & b & 0 & 0 \\
	-b & a & 0 & 0 \\
	0 & 0 & a & -b \\
	0 & 0 & b & a
	\end{matrix}\right),
\label{eq:corep_di_2D_m1} \\
D^2_{m^2_2}(R) &= \left(\begin{matrix} 
	a & b & 0 & 0 \\
	b & -a & 0 & 0 \\
	0 & 0 & a & -b \\
	0 & 0 & -b & -a
	\end{matrix}\right).
\label{eq:corep_di_2D_m2}
\end{align}
From Eq.~\eqref{eq:corep_di_antiunitary}, we obtain the corep of $A$ as
\begin{equation}\label{eq:corep_2D_A}
	D^2(A)= \left(\begin{matrix}
	0 & \bbone_2 \\
	\bbone_2 & 0
	\end{matrix}\right). 
\end{equation}
Next, we diagonalize $D^2(A)$ such that each main-diagonal block matrix is given by Eq.~\eqref{eq:mat_rep_chi_0},
\begin{equation}\label{eq:corep_di_2D_A_diag}
D'^2(A)=(U^2)^{-1}D^2(A)U^2=\left(\begin{matrix} 
1 & 0 & 0 & 0 \\
0 & -1 & 0 & 0 \\
0 & 0 & 1 & 0 \\
0 & 0 & 0 & -1
\end{matrix}\right), 
\end{equation}
with the orthogonal transformation
\begin{equation}
	U^2=\frac{1}{\sqrt{2}}\left(\begin{matrix}
	-1 & 0 & 0 & 1 \\
	0 & -1 & -1 & 0 \\
	-1 & 0 & 0 & -1 \\
	0 & 1 & -1 & 0
	\end{matrix}\right). 
\end{equation}
Using Eq.~\eqref{eq:coreps_uni_trafo_uni}, we find that the unitary elements for the corep $D'^2$ are of the form 
\begin{align}
D'^2_{m^2_1}(R) &= \left(\begin{matrix} 
	a & b & 0 & 0 \\
	-b & a & 0 & 0 \\
	0 & 0 & a & b \\
	0 & 0 & -b & a
	\end{matrix}\right),
\label{eq:corep_di_2D_m1_diag} \\
D'^2_{m^2_2}(R) &= \left(\begin{matrix} 
	a & b & 0 & 0 \\
	b & -a & 0 & 0 \\
	0 & 0 & -a & -b \\
	0 & 0 & -b & a
	\end{matrix}\right).
\label{eq:corep_di_2D_m2_diag}
\end{align}
Thus, the orthogonal transformation $U^2$ diagonalizes $D^2(A)$ and leaves coreps of the unitary elements block diagonal, as seen from Eqs.~\eqref{eq:corep_di_2D_A_diag}, \eqref{eq:corep_di_2D_m1_diag}, and \eqref{eq:corep_di_2D_m2_diag}. Furthermore, as shown by Eqs.~\eqref{eq:corep_di_2D_m1}, \eqref{eq:corep_di_2D_m2}, \eqref{eq:corep_di_2D_m1_diag}, and \eqref{eq:corep_di_2D_m2_diag}, for each $i$ the top-left block matrices of $D^2_{m^2_i}(R)$ and $D'^2_{m^2_i}(R)$ are identical, whereas the bottom-right block matrices have the same trace but are not identical. The remaining coreps of the antiunitary elements $B\in A\bm{H}\setminus A$ of $D'^2$ are calculated using Eqs.~\eqref{eq:calc_rules_coreps_3}, \eqref{eq:corep_di_2D_A_diag}, \eqref{eq:corep_di_2D_m1_diag}, and \eqref{eq:corep_di_2D_m2_diag},
\begin{align}
D'^2_{m^2_1}(B) &= \left(\begin{matrix} 
a & b & 0 & 0 \\
b & -a & 0 & 0 \\
0 & 0 & a & b \\
0 & 0 & b & -a
\end{matrix}\right),
\label{eq:corep_di_2D_antiuni_m1_diag} \\
D'^2_{m^2_2}(B) &= \left(\begin{matrix} 
a & b & 0 & 0 \\
-b & a & 0 & 0 \\
0 & 0 & -a & -b \\
0 & 0 & b & -a
\end{matrix}\right).
\label{eq:corep_di_2D_antiuni_m2_diag}
\end{align}
Hence, by inspection of the coreps in Eqs.~\eqref{eq:corep_di_2D_m1_diag}--\eqref{eq:corep_di_2D_antiuni_m2_diag}, we find that the corep $D'^2$ is reducible into two two-dimensional real coreps,
\begin{equation}\label{eq:2D_corep_real_coreps}
	D'^2=d^2 \oplus d'^2.  
\end{equation}
In general, $d^2$ and $d'^2$ are different real coreps. However, as seen from Eqs.~\eqref{eq:corep_di_2D_antiuni_m1_diag} and \eqref{eq:corep_di_2D_antiuni_m2_diag}, the real coreps $d^2$ and $d'^2$ are identical if there are no coreps of the form $D'^2_{m^2_2}$, i.e., if the representation $\Delta^2$ lacks matrix representatives of the form $m^2_2$ for the elements of the halving subgroup $\bm{H}$. 

It remains to investigate the basis functions. As for the one-dimensional coreps of dichromatic point groups, we differentiate between basis functions of $\Delta^2$ that are even with respect to $\Theta$, $\psi_2^e=(\psi_{21}^e,\psi_{22}^e)$, and basis functions that are odd, $\psi_2^o=(\psi_{21}^o,\psi_{22}^o)$. The basis functions of $D^2$ are calculated as
\begin{align}
	\gamma_2^e &= (\psi_2^e,\hat{P}_{\Theta R'}\psi_2^e)=(\psi_{21}^e,\psi_{22}^e,\psi_{21}^e,-\psi_{22}^e),\\
	\gamma_2^o &= (\psi_2^o,\hat{P}_{\Theta R'}\psi_2^o)=(\psi_{21}^o,\psi_{22}^o,-\psi_{21}^o,\psi_{22}^o). 
\end{align}
Consequently, the basis functions of $D'_2$ are
\begin{align}
	\gamma'^e_2=\gamma_2^e U^2 = -\sqrt{2}\, (\psi_{21}^e,\psi_{22}^e,0,0),\\
	\gamma'^o_2=\gamma_2^o U^2 = \sqrt{2}\, (0,0,-\psi_{22}^o,\psi_{21}^o). 
\end{align}
The first two components of $\gamma'^e_2$ are already proportional to $\psi_2^e$. By performing another orthogonal transformation we can bring the last two components of $\gamma'^o_2$ into the form $\psi_2^o = (\psi_{21}^o,\psi_{22}^o)$, where the transformation keeps $D'^2$ still block diagonal. For the rest of this Appendix, we omit mentioning this step. We conclude that $\psi_2^e$ is a basis function of $d_2$, and $\psi_2^o$ is a basis function of $d'_2$. 

We now turn to the case with $\chi(\Delta^2(R'))=\pm 2$. Then the matrix representative $\Delta^2(R')$ is of the form $m^2_1$ in Eq.~\eqref{eq:matrix_rep_two_dim} with $a=\pm 1$ and $b=0$. Thus, $\Delta^2(R')$ is the identity matrix, up to a sign. In the following, we consider only the case with
\begin{equation}
	\Delta^2(R')= \left(\begin{matrix}
	1 & 0 \\
	0 & 1
	\end{matrix}\right).
\end{equation} 
The case of the opposite sign is analogous. Since $\Delta^2(R')$ is the identity matrix the coreps of the unitary elements are given by
\begin{equation}\label{eq:corep_di_2D_unit}
D^2_{m^2_i}(R)=\left(\begin{matrix}
m^2_i & 0 \\
0 & \Delta^2(R')^{-1}m^2_i\Delta^2(R')
\end{matrix}\right)=\left(\begin{matrix}
m^2_i & 0 \\
0 & m^2_i
\end{matrix}\right)
\end{equation} 
and the corep of $A$ is still given by Eq.~\eqref{eq:corep_2D_A}. In order to diagonalize $D^2(A)$, we rearrange the columns of $U^2$ in the form
\begin{equation}
	\tilde{U}^2=\frac{1}{\sqrt{2}}\left(\begin{matrix}
	-1 & 0 & 1 & 0 \\
	0 & -1 & 0 & -1 \\
	-1 & 0 & -1 & 0 \\
	0 & -1 & 0 & 1
	\end{matrix}\right). 
\end{equation}
$\tilde{U}^2$ diagonalizes the corep of $A$ in such a way that each main-diagonal block matrix is proportional to the identity matrix,
\begin{equation}\label{eq:corep_di_2D_A_diag_unit}
	\tilde{D}'^2(A)=({ \tilde{U}^2})^{-1}D^2(A){\tilde{U}^2}=\left(\begin{matrix} 
	1 & 0 & 0 & 0 \\
	0 & 1 & 0 & 0 \\
	0 & 0 & -1 & 0 \\
	0 & 0 & 0 & -1
	\end{matrix}\right), 
\end{equation}
and the coreps of the unitary elements again remain block diagonal,
\begin{align}
\tilde{D}'^2_{m^2_1}(R) &= \left(\begin{matrix} 
	a & b & 0 & 0 \\
	-b & a & 0 & 0 \\
	0 & 0 & a & -b \\
	0 & 0 & b & a
	\end{matrix}\right),
\label{eq:corep_di_2D_m1_diag_unit} \\
\tilde{D}'^2_{m^2_2}(R) &= \left(\begin{matrix} 
	a & b & 0 & 0 \\
	b & -a & 0 & 0 \\
	0 & 0 & a & -b \\
	0 & 0 & -b & -a
	\end{matrix}\right).
\label{eq:corep_di_2D_m2_diag_unit}
\end{align}
Thus, the corep $\tilde{D}'^2$ decomposes into two different real coreps,
\begin{equation}
	\tilde{D}'^2=d^2_+\oplus d^2_-,
\end{equation} 
where $d^2_+$ and $d^2_-$ are even and odd with respect to $A$, respectively. The basis functions of $D^2$ are calculated as 
\begin{align}
\gamma_2^e &= (\psi_2^e,\hat{P}_{\Theta R'}\psi_2^e)=(\psi_{21}^e,\psi_{22}^e,\psi_{21}^e,\psi_{22}^e),\\
\gamma_2^o &= (\psi_2^o,\hat{P}_{\Theta R'}\psi_2^o)=(\psi_{21}^o,\psi_{22}^o,-\psi_{21}^o,-\psi_{22}^o). 
\end{align}
Thus, for the basis functions of $\tilde{D}'^2$ we obtain
\begin{align}
\tilde{\gamma}'^e_2=\gamma_2^e { \tilde{U}^2} = -\sqrt{2}\, (\psi_{21}^e,\psi_{22}^e,0,0),\\
\tilde{\gamma}'^o_2=\gamma_2^o { \tilde{U}^2} = \sqrt{2}\, (0,0,\psi_{21}^o,-\psi_{22}^o), 
\end{align}
which shows that $\psi_2^e$ and $\psi_2^o$ belong to the real coreps that are even and odd with respect to $A$, respectively. 

In analogy to the one-dimensional real coreps, the labeling of a two-dimensional real corep of a dichromatic point group $\bm{G}(\bm{H})$ can be chosen to be one of the two-dimensional representations of the monochromatic point group $\bm{G}$. In order to identify a two-dimensional real corep $\tilde{d}^2$ with the two-dimensional real representation $\Delta^2$, we have to distinguish the two cases $\chi(\Delta^2(R'))=0$ and $\chi(\Delta^2(R'))=\pm 2$. For the latter case, a real corep $\tilde{d}^2$ and a representation $\Delta^2$ are identified if all characters of the unitary elements in $\bm{H}$ coincide and
\begin{equation}\label{eq:label_2D_corep_rep_A}
	\chi(\Delta^2(R'))=\chi(\tilde{d}^2(\Theta R'))
\end{equation}
holds. However, for the case with $\chi(\Delta^2(R'))=0$, we cannot rely on Eq.~\eqref{eq:label_2D_corep_rep_A} to identify a representation with a real corep for the following reason. The corep $D'^2$ is reducible into two two-dimensional real coreps $d^2$ and $d'^2$, which can be different; see the discussion of Eq.~\eqref{eq:2D_corep_real_coreps}. For the case $\chi(\Delta^2(R'))=0$, both have zero character for $A=\Theta R'$, as reflected by Eq.~\eqref{eq:corep_di_2D_A_diag}.
Note that we can write every monochromatic point group as 
\begin{equation}
	\bm{G}=\bm{H}+R'\bm{H}.
\end{equation}
If all matrix representatives $\Delta^2$ of the halving subgroup $\bm{H}$ are of the form $m_1^2$ in Eq.~\eqref{eq:matrix_rep_two_dim}, then all the matrix representatives of $R'\bm{H}$ are of the form
\begin{equation}
	\Delta^2(R') m_1^2 = m_2^2,
\end{equation}
where we have used Eq.~\eqref{eq:mat_rep_chi_0}. Thus, the characters of $\Delta^2$ of the conjugacy classes of the elements $\tilde{R}\in R'\bm{H}$ vanish and $\Delta^2$ is thus determined by the characters of $R\in\bm{H}$ alone. Consequently, the representation $\Delta^2$ is identified with that real corep $\tilde{d}^2$ that shares the same characters for the conjugacy classes of $\bm{H}$.

On the other hand, if there are matrix representatives of the form $m_2^2$ for some elements of the halving subgroup $\bm{H}$, then there are matrix representatives of the form 
\begin{equation}
	\Delta^2(R')m_2^2=m_1^2
\end{equation}
for some elements of the left coset $R'\bm{H}$ so that some characters can be nonzero. In fact, one can verify two statements for a two-dimensional representation that has matrix representatives of the form $m_2^2$ for some elements of the halving subgroup $\bm{H}$. First, there are nonzero characters for some conjugacy classes of $R'\bm{H}$ \cite{Altmann_Herzig_1994}. Second, there exists a second two-dimensional representation whose characters for the conjugacy classes of the elements of $\bm{H}$ are identical to the characters in the first representation, whereas the nonzero characters of the conjugacy classes of $R'\bm{H}$ are inverted in sign \cite{Altmann_Herzig_1994}. In this case, we identify $\Delta^2$ with that real corep $\tilde{d}^2$ for which the characters of the elements in $\bm{H}$ of $\Delta^2$ and $\tilde{d}^2$ are identical and for which
\begin{equation}
	\chi(\Delta^2(\tilde{R}))=\chi(\tilde{d}^2(\Theta\tilde{R}))\neq 0
\end{equation}
holds for one $\tilde{R}\in\bm{G}-\bm{H}$ with $\Theta\tilde{R}\in \Theta(\bm{G}-\bm{H})$. Since all $\tilde R$ are elements of $R'\bm{H}$ the previous argument shows that such a $\tilde{R}$ has to exist.

\begin{table*}
	\begin{center}
		\caption{Character table of the dichromatic point group $D_4(C_4)$ with basis functions up to second order in momentum and spin basis functions. The lowest-order basis function of $A_2$ is $k_x k_y (k_x^2-k_y^2)$.}
		\begin{tabular}{c c  c  c  c  c c }
			\hline\hline
			$D_4(C_4)$ & $e$  & $2C_{4}$ & $C_2$ & $2 \Theta C'_{2}$ & $2 \Theta C''_{2}$ & basis functions\\\hline
			\rule{0pt}{3ex}
			$A_1$ & $1$ & $1$ & $1$ & $1$ & $1$ &$\sigma_0$, $\sigma_z$, $1$, $k_z$, $k_x^2+k_y^2$, $k_z^2$ \\
			\rule{0pt}{3ex}
			$A_2$ & $1$ & $1$ & $1$ & $-1$ & $-1$ & \\
			\rule{0pt}{3ex}
			$B_1$ & $1$ & $-1$ & $1$ & $1$ & $-1$ & $k_x^2-k_y^2$ \\
			\rule{0pt}{3ex}
			$B_2$ & $1$ & $-1$ & $1$ & $-1$ & $1$ & $k_x k_y$ \\
			\rule{0pt}{3ex}
			$E$ & $2$ & $0$ & $-2$ & $0$ & $0$ & $\{\sigma_x,\sigma_y\}$, $\{k_x,k_y\}$,
			$\{k_z k_x ,k_z k_y\}$, $\{k_x k_z^2, k_y k_z^2\}$,
			$\{k_x (k_x^2 - 3 k_y^2), k_y (k_y^2-3 k_x^2)\}$ \\
			\hline\hline
		\end{tabular}
		\label{table:char_tab_D4_C4}
	\end{center} 	
\end{table*}

In the following, we choose this labeling and discuss two examples of two-dimensional real coreps. First, we consider the monochromatic point group $D_4$, which has the halving subgroup $\bm{H}=C_4$. $D_4$ has one two-dimensional $E$ representation. Ref.~\cite{Bradley_Cracknell_1972} shows that all matrix representatives of the elements of $\bm{H}$ are of the form $m^2_1$ in Eq.~\eqref{eq:matrix_rep_two_dim}. The elements of the set $\bm{G}-\bm{H}$ correspond to twofold rotations with trace zero. Thus, the corep derived from $E$ for the dichromatic point group $D_4(C_4)$ reduces to twice the two-dimensional real corep that we also label $E$. The character table of $D_4(C_4)$ is given in Table~\ref{table:char_tab_D4_C4}. 

As a second example, we discuss the real coreps of the dichromatic point group $C_{6v}(C_{3v})$. Here, we consider the corep derived from the two-dimensional $E_1$ representation of $C_{6v}$. The set $C_{6v}-C_{3v}$ contains four group elements that correspond to twofold symmetries, namely one rotation $C_2$ and three mirror planes $\sigma_{di}$ with $i=1,2,3$. First, let us choose $A=\Theta C_2$ where $\Delta^{E_1}(C_2)=-2$. Then, as shown by the analysis above, we find that there are two distinct two-dimensional real coreps, with the labels $E_1$ and $E_2$. Instead of choosing $A=\Theta C_2$, we could also choose $A=\Theta \sigma_{d1}$. This choice leads to the same conclusions since we also find the two two-dimensional real coreps $E_1$ and $E_2$ because $\Delta^{E_1}(\sigma_{d1})=0$ and the matrix representatives of two of the three vertical mirror planes $\sigma_v$ in $C_{3v}$ are of the form $m^2_2$ \cite{Bradley_Cracknell_1972}. The character table of the two two-dimensional real coreps of $C_{6v}(C_{3v})$ is given in Table \ref{table:char_tab_C6v_C3v}.   

\begin{table}
	\begin{center}
		\caption{Character table of the two-dimensional real coreps of the dichromatic point group $C_{6v}(C_{3v})$.}
		\begin{tabular}{c c  c  c  c  c c }
			\hline\hline
			$C_{6v}(C_{3v})$ & $e$  & $2C_{3}$ & $3\sigma_v$ & $2 \Theta C_{6}$ & $\Theta C_{2}$ & $3\Theta\sigma_d$\\\hline
			\rule{0pt}{3ex}
			$E_1$ & $2$ & $-1$ & $0$ & $1$ & $-2$ & $0$ \\
			
			\rule{0pt}{3ex}
				
			$E_2$ & $2$ & $-1$ & $0$ & $-1$ & $2$ & $0$ \\
			\hline\hline
		\end{tabular}
		\label{table:char_tab_C6v_C3v}
	\end{center} 	
\end{table} 

\subsubsection{Coreps of dichromatic point groups derived from three-dimensional representations}

The highest-dimensional representations that occur for the 58 dichromatic point groups are three-dimensional, i.e., $T$, representations. In this section, we derive the real corep from a three-dimensional representation $\Delta^3$ for the case that the antiunitary element $A=\Theta R'$ can be chosen such that $R'^2=e$. For a three-dimensional representation, the character of $R'$ can take four different values that are given by $\chi(\Delta^3(R'))=\pm 3$ and $\chi(\Delta^3(R'))=\pm 1$.

First, we consider the case $\chi(\Delta^3(R'))= \pm 3$. Then, $R'$ corresponds to the inversion $i$ so that the antiunitary element is given by $A=\Theta R'=\Theta i$. Up to a sign, real matrix representatives of a three-dimensional representation can be written as \cite{Bradley_Cracknell_1972} 
\begin{align}
m^3_1 &= \left(\begin{matrix}
	1 & 0 & 0 \\
	0 & 1 & 0 \\
	0 & 0 & 1 \\
	\end{matrix}\right),& m^3_2 &= \left(\begin{matrix}
	1 & 0 & 0 \\
	0 & -1 & 0 \\
	0 & 0 & -1 \\
	\end{matrix}\right), \\
 m^3_3 &= \left(\begin{matrix}
	-1 & 0 & 0 \\
	0 & 1 & 0 \\
	0 & 0 & -1 \\
	\end{matrix}\right),& m^3_4 &= \left(\begin{matrix}
	-1 & 0 & 0 \\
	0 & -1 & 0 \\
	0 & 0 & 1 \\
	\end{matrix}\right), \\
m^3_5 &= \left(\begin{matrix}
	0 & 1 & 0 \\
	0 & 0 & 1 \\
	1 & 0 & 0 \\
	\end{matrix}\right),& m^3_6 &= \left(\begin{matrix}
	0 & 0 & 1 \\
	1 & 0 & 0 \\
	0 & 1 & 0 \\
	\end{matrix}\right), \\
m^3_7 &= \left(\begin{matrix}
	0 & 1 & 0 \\
	1 & 0 & 0 \\
	0 & 0 & -1 \\
	\end{matrix}\right), &&
\label{eq:3D_rep_m7}
\end{align}  
or as products of these matrices. In the following, we assume that the matrix representative of $R'=i$ is
\begin{equation}\label{eq:rep_3D_Rp}
\Delta^3(R')=m_1^3,  
\end{equation}
i.e., we consider the case $\chi(\Delta^3(R'))=3$. Using Eq.~\eqref{eq:corep_di_antiunitary}, we obtain the corep of the antiunitary element $A$ as
\begin{equation}
	D^3(A)= \left(\begin{matrix}
0 & \bbone_3 \\
\bbone_3 & 0
\end{matrix}\right). 
\end{equation}
For the unitary elements, we use Eq.~\eqref{eq:corep_di_2D}, where we just have to replace the two-dimensional representation with the three-dimensional one. Then, coreps of the unitary elements are found as
\begin{equation}\label{eq:uni_elements_3D_corep_case_1}
	D^3_{m^3_i}(R)=\left(\begin{matrix}
	m^3_i & 0 \\
	0 & m^3_i
	\end{matrix}\right). 
\end{equation}
The orthogonal transformation
\begin{equation}\label{eq:uni_trafo_3D_reps_case_1}
	U^3=\frac{1}{\sqrt{2}}\left(\begin{matrix} 
	0 & 0 & -1 & 0 & 0 & 1 \\
	0 & -1 & 0 & 0 & 1 & 0 \\
	-1 & 0 & 0 & 1 & 0 & 0 \\
	0 & 0 & 1 & 0 & 0 & 1 \\
	0 & 1 & 0 & 0 & 1 & 0\\
	1 & 0 & 0 & 1 & 0 & 0\\
	\end{matrix}\right)
\end{equation} 
diagonalizes $D^3(A)$ such that one of the two main-diagonal blocks is given by Eq.~\eqref{eq:rep_3D_Rp},
\begin{equation}
	D'^3(A)=(U^3)^{-1} D^3(A) {U^3} =\left(\begin{matrix}
	-\bbone_3 & 0 \\
	0 & \bbone_3
	\end{matrix}\right). 
\end{equation}
For the coreps of the unitary elements in Eq.~\eqref{eq:uni_elements_3D_corep_case_1}, one can check that each matrix representative is left block diagonal. Thus, the corep $D^3$ derived from the three-dimensional representation with $\Delta^3(R')=m_1^3$ decomposes into two real coreps 
\begin{equation}
	D'^3=d^3_- \oplus d^3_+,
\end{equation}
where $d^3_-$ and $d^3_+$ are odd and even with respect to $A$, respectively. For the basis functions, we again distinguish between basis functions that are even with respect to $\Theta$, $\psi_3^e=(\psi_{31}^e,\psi_{32}^e,\psi_{33}^e)$, and basis functions that are odd, $\psi_3^o=(\psi_{31}^o,\psi_{32}^o,\psi_{33}^o)$. Thus, the basis functions of $D^3$ are given by
\begin{align}
\gamma_3^e &= (\psi_3^e,\hat{P}_{\Theta R'}\psi_3^e)=(\psi_{31}^e,\psi_{32}^e,\psi_{33}^e,\psi_{31}^e,\psi_{32}^e,\psi_{33}^e),\\
\gamma_3^o &= (\psi_3^o,\hat{P}_{\Theta R'}\psi_3^o)=(\psi_{31}^o,\psi_{32}^o,\psi_{33}^o,-\psi_{31}^o,-\psi_{32}^o,-\psi_{33}^o). 
\end{align}
Hence, we find for the basis functions of $D'^3$ that
\begin{align}\label{eq:basis_fct_3D_corep}
\gamma'^e_3 &= \gamma_3^e { U^3} = \sqrt{2}\, (0,0,0,\psi_{33}^e,\psi_{32}^e,\psi_{31}^e),\\
\gamma'^o_3 &= \gamma_3^o {U^3} = -\sqrt{2}\, (\psi_{33}^o,\psi_{32}^o,\psi_{31}^o,0,0,0). 
\end{align}
Consequently, $\psi_3^e$ and $\psi_3^o$ are basis functions of $d^3_+$ and $d^3_-$, respectively. In the following, we discuss this result in more detail. Since we have derived the corep $D^3$ from a three-dimensional representation with $\chi(\Delta^3(R'))=\chi(\Delta^3(i))=3$, it is clear that $\Delta^3$ is a $T_g$ representation, e.g., the $T_{1g}$ representation of $O_h$. For a $g$ representation, the basis functions $\psi_3^e$ that are even under $A$ are of even order in momentum components, whereas the odd basis functions $\psi_3^o$ are the Pauli matrices. Thus, Eq.~\eqref{eq:basis_fct_3D_corep} shows that the even-in-momentum basis functions and the Pauli matrices are in different real coreps. In analogy to the just shown procedure, one can derive the real coreps from a $u$ representation with $\chi(\Delta^3(R'))=\chi(\Delta^3(i))=-3$. Since the basis functions of a $u$ representation are odd in momentum, they are even with respect to the product $\Theta i$. Thus, an odd-in-momentum basis function of a three-dimensional representation belongs to a real corep that is even with respect to $A$. This is Kramers' theorem at the level of real coreps: Since for magnetic point groups that contain $\Theta i$ the Pauli matrices and momentum basis functions are never in the same real corep the bands are doubly degenerate. 

Now, we consider the case $\chi(\Delta^3(R'))= - 1$, where we choose $m^3_7$ in Eq.~\eqref{eq:3D_rep_m7} as the matrix representative of $R'$, i.e.,
\begin{equation}\label{eq:3D_rep_case_2}
	\Delta^3(R') = m^3_7. 
\end{equation}
The coreps of the unitary elements are derived using Eq.~\eqref{eq:corep_di_2D}, where we replace $m_i^2$ by $m_i^3$ and use Eq.~\eqref{eq:3D_rep_case_2} for the matrix representative $\Delta^3(R')$. Next, we apply the orthogonal transformation
\begin{equation}\label{eq:uni_trafo_3D_reps_case_2}
\tilde{U}^3=\frac{1}{\sqrt{2}}\left(\begin{matrix} 
1 & 0 & 0 & -1 & 0 & 0 \\
0 & 1 & 0 & 0 & -1 & 0 \\
0 & 0 & -1 & 0 & 0 &1 \\
0 & -1 & 0 & 0 & -1 & 0 \\
-1 & 0 & 0 & -1 & 0 & 0\\
0 & 0 & -1 & 0 & 0 & -1\\
\end{matrix}\right)
\end{equation} 
to the corep $D^3$. As one can verify, $ \tilde{U}^3$ leaves the coreps of the unitary elements block-di\-a\-go\-nal. For the corep of the antiunitary element $A$, we find 
\begin{equation}
	{\tilde{D}}'^3(A)=({\tilde{U}^3})^{-1} D^3(A) {\tilde{U}^3}=\left(\begin{matrix} 
	0 & -1 & 0 & 0 & 0 & 0 \\
	-1 & 0 & 0 & 0 & 0 & 0 \\
	0 & 0 & 1 & 0 & 0 & 0 \\
	0 & 0 & 0 & 0 & 1 & 0 \\
	0 & 0 & 0 & 1 & 0 & 0\\
	0 & 0 & 0 & 0 & 0 & -1\\
	\end{matrix}\right),
\end{equation}
where the bottom-right block matrix is identical to $\Delta(R')=m^3_7$. Thus, the corep $D'^3$ decomposes into two different real coreps $d'^3$ and $d^3$,
\begin{equation}
	\tilde{D}'^3 = d'^3 \oplus d^3, 
\end{equation}
where $\chi(d'^3(A))=1$ and $\chi(d^3(A))=-1$. For the basis functions, we find that 
\begin{equation}
\gamma_3^e = (\psi_3^e,\hat{P}_{\Theta R'}\psi_3^e)=(\psi_{31}^e,\psi_{32}^e,\psi_{33}^e,\psi_{32}^e,\psi_{31}^e,-\psi_{33}^e)
\end{equation}
and
\begin{equation}
\gamma_3^o = (\psi_3^o,\hat{P}_{\Theta R'}\psi_3^o)=(\psi_{31}^o,\psi_{32}^o,\psi_{33}^o,-\psi_{32}^o,-\psi_{31}^o,\psi_{33}^o)
\end{equation}
transform as
\begin{align}\label{eq:basis_fct_3D_corep_2}
\tilde{\gamma}'^e_3 &= \gamma_3^e \tilde{U}^3 = \sqrt{2} (0,0,0,-\psi_{31}^e,-\psi_{32}^e,\psi_{33}^e),\\
\tilde{\gamma}'^o_3 &= \gamma_3^o \tilde{U}^3 = \sqrt{2} (\psi_{31}^o,\psi_{32}^o,-\psi_{33}^o,0,0,0).
\end{align}
Thus, $\psi^e_3$ and $\psi^o_3$ are basis functions of the real coreps with $\chi(d^3(A))=-1$ and $\chi(d'^3(A))=1$, respectively. 

In analogy to one-dimensional real coreps, the labeling of three-dimensional real coreps can be chosen to be the labeling of the three-dimensional representations from which they are derived. In particular, one identifies the three-dimensional representation with that three-dimensional real corep $\tilde{d}^3$ for which the characters of the unitary elements in $\bm{H}$ are identical and for which
\begin{equation}
	\chi(\Delta^3(R'))=\chi(\tilde{d}^3(\Theta R'))
\end{equation}
holds.

\subsubsection{Corepresentations of the dichromatic point group $S_4(C_2)$}

\begin{table}
	\begin{center}
		\caption{Character table of the monochromatic point group $S_4$ with basis functions up to second order in momentum and spin basis functions.}
		\begin{tabular}{c c  c  c   c }
			\hline\hline
			$S_4$ & $e$  & $C_2$ & $2 S_4$ &  basis functions\\\hline
			\rule{0pt}{3ex}
			$A$ & $1$ & $1$ & $1$ & $\sigma_z,\, 1,\, k_x^2+k_y^2,\, k_z^2 $ \\
			\rule{0pt}{3ex}
			$B$ & $1$ & $1$ & $-1$ & $k_z,\, k_x^2 - k_y^2,\, k_x k_y$ \\
			\rule{0pt}{3ex}
			$E$ & $2$ & $-2$ & $0$ & $\{\sigma_x,\sigma_y\},\,\{k_x,k_y\},\,\{k_x k_z, k_y k_z\}$\\
			\hline\hline
		\end{tabular}
		\label{table:char_tab_S4}
	\end{center} 	
\end{table} 

Finally, we consider the dichromatic point group $S_4(C_2)$ as an example for which no antiunitary element $A=\Theta R'$ with $R'^2=e$ exists. For $S_4$, the character table is given in Table \ref{table:char_tab_S4}. First, we investigate the coreps derived from the $B$ representations. For the antiunitary element, we choose $A=\Theta S_4^+$. Since $(S_4^+)^{-1} R S_4^+$ is an element of the conjugacy class of $R$, there is nothing further to calculate for the coreps of the unitary elements $R$: for the two unitary elements $R = e, C_2$, the corep is the identity matrix
\begin{equation}\label{eq:corep_1D_S4_C4_uni}
	D^B(R)= \left(\begin{matrix}
	1 & 0 \\
	0 & 1
	\end{matrix}\right). 
\end{equation}
Using Eq.~\eqref{eq:def_corep_antiuni} and that 
\begin{equation}
	\Delta^B((\Theta S_4^+)(\Theta S_4^+))=\Delta^B(\Theta^2 (S_4^+)^2)=\Delta^B(\tilde{e} C_2)=1,
\end{equation}
the corep of $A$ is calculated as
\begin{equation}\label{eq:corep_1D_S4_C4_A}
D^B(A)= \left(\begin{matrix}
0 & 1 \\
1 & 0
\end{matrix}\right). 
\end{equation}
Thus, the matrix representatives of $D^1$ are diagonalized by $U^1$ in Eq.~\eqref{eq:uni_trafo_U1} and we obtain
\begin{equation}
    D'^B(R)= \left(\begin{matrix}
	1 & 0 \\
	0 & 1
	\end{matrix}\right),\quad D'^B(B)= \left(\begin{matrix}
	-1 & 0 \\
	0 & 1
	\end{matrix}\right). 
\end{equation}
Hence, $D'^B$ is reducible to two one-dimensional real coreps,
\begin{equation}
	 D'^B=B\oplus A.
\end{equation}
Next, we have a look at the basis functions, where we separately consider basis functions of the representation $B$ that are even $ (\psi_B^e)$ and odd $(\psi_B^o)$ with respect to $\Theta$. For an even basis function, the basis function $\gamma_B^e$ of the corep $D$ is calculated as
\begin{equation}
\gamma_B^e=(\psi_B^e,\hat{P}_{\Theta S_4^+} \psi_B^e)=(\psi_B^e,-\psi_B^e),
\end{equation} 
where we have used that $\psi_B^e$ is odd with respect to $S_4^+$ since we are deriving the corep from the $B$ representation. For $\psi_B^o$, we analogously find
\begin{equation}
\gamma_B^o=(\psi_B^o,\hat{P}_{\Theta S_4^+} \psi_B^o)=(\psi_B^o,\psi_B^o). 
\end{equation} 
Thus, the basis functions of $D'$ are computed as 
\begin{align}
	\gamma'^e_B &= -\sqrt{2}\, (\psi_B^e,0), \\
	\gamma'^o_B &= \sqrt{2}\, (0,\psi_B^o). 
\end{align}
Consequently, $\psi_B^e$ and $\psi_B^o$ are basis functions of the $B$ corep and the $A$ corep, respectively. The derivation of the real coreps from the trivial $A$ representation works analogously. 

\begin{table}
	\begin{center}
		\caption{Character table of the dichromatic point group $S_4(C_2)$ with basis functions up to second order in momentum and spin basis functions.}
		\begin{tabular}{c c  c  c   c }
			\hline\hline
			$S_4(C_2)$ & $e$  & $C_2$ & $2 \Theta S_4$ &  basis functions\\\hline
			\rule{0pt}{3ex}
			$A$ & $1$ & $1$ & $1$ & $1,\,k_z,\, k_x^2+k_y^2,\, k_z^2 $\\
			\rule{0pt}{3ex}
			$B$ & $1$ & $1$ & $-1$ & $\sigma_z,\, k_x^2 - k_y^2,\, k_x k_y$ \\
			\rule{0pt}{3ex}
			$E$ & $2$ & $-2$ & $0$ & $\{\sigma_x,\sigma_y\},\,\{k_x,k_y\},\,\{k_x k_z, k_y k_z\}$ \\
			\hline\hline
		\end{tabular}
		\label{table:char_tab_S4_C2}
	\end{center} 	
\end{table} 

Next, we derive coreps of $S_4(C_2)$ starting from the $E$ representation of $S_4$. The matrix representatives of $S_4$ are given in Ref.~\cite{Bradley_Cracknell_1972},
\begin{align}
\Delta^E(e) &= \left(\begin{matrix}
	1 & 0 \\
	0 & 1
	\end{matrix}\right),\quad \Delta^E(C_{2z})= -\left(\begin{matrix}
	1 & 0 \\
	0 & 1
	\end{matrix}\right), \\
\Delta^E(S_{4}^+) &= \left(\begin{matrix}
	0 & -1 \\
	1 & 0
	\end{matrix}\right),\quad \Delta^E(S_{4}^-)= \left(\begin{matrix}
	0 & 1 \\
	-1 & 0
	\end{matrix}\right). 
\label{eq:matrix_reps_S4_2}
\end{align}
Then, the coreps derived from $E$ are computed as
\begin{align}
D^E(e) &= \left(\begin{matrix}
\bbone_2 & 0 \\
0 & \bbone_2
\end{matrix}\right),& D^E(C_{2z}) &= -\left(\begin{matrix}
\bbone_2 & 0 \\
0 & \bbone_2
\end{matrix}\right), \\
D^E(\Theta S_{4}^+) &= \left(\begin{matrix}
0 & -\bbone_2 \\
\bbone_2 & 0
\end{matrix}\right),&
D^E(\Theta S_4^-) &= \left(\begin{matrix}
	0 & \bbone_2 \\
	-\bbone_2 & 0
	\end{matrix}\right),
\end{align}
where the corep for $\Theta S_4^-$ follows from the product rules of the coreps. Since $D^E(\Theta S_4^+)$ is a real skew-symmetric matrix it cannot be orthogonally diagonalized. However, we can block-diagonalize $D^E(\Theta S_4^+)$ in two steps. The first step is to apply the orthogonal transformation 
\begin{equation}
	U_E=\left(\begin{matrix} 
	-1 & 0 & 0 & 0 \\
	0 & -1 & 0 & 0 \\
	0 & 0 & 0 & -1 \\
	0 & 0 & 1 & 0
	\end{matrix}\right),
\end{equation}
which yields
\begin{equation}\label{eq:corep0_theta_s4}
	D^E_0(\Theta S_4^+)=U_E^{-1} D^2(\Theta S_{4}^+) U_E=\left(\begin{matrix} 
	0 & 0 & 0 & -1 \\
	0 & 0 & 1 & 0 \\
	0 & -1 & 0 & 0 \\
	1 & 0 & 0 & 0
	\end{matrix}\right)
\end{equation}
so that each off-diagonal block matrix is identical to $\Delta^E(S_{4}^+)$ in Eq.~\eqref{eq:matrix_reps_S4_2}. Since the coreps of $e$ and $C_{2z}$ are proportional to the identity matrix, $U_0$ leaves them diagonal. Note that we can rewrite Eq.~\eqref{eq:corep0_theta_s4} as
\begin{equation}
	D^E_0(\Theta S_4^+)=\sigma_x\otimes (-i \sigma_y).
\end{equation}
The second step is to apply the orthogonal transformation
\begin{equation}
\tilde{U}_E = U^1\otimes \bbone_2,
\end{equation}
which block-diagonalizes the corep $D^E_0$,
\begin{align}
D'^E(e) &= \left(\begin{matrix}
\bbone_2 & 0 \\
0 & \bbone_2
\end{matrix}\right), \\
D'^E(C_{2z}) &= -\left(\begin{matrix}
\bbone_2 & 0 \\
0 & \bbone_2
\end{matrix}\right), \\
D'^E(\Theta S_{4}^+) &= \left(\begin{matrix} 
0 & -1 & 0 & 0 \\
1 & 0 & 0 & 0 \\
0 & 0 & 0 & 1 \\
0 & 0 & -1 & 0
\end{matrix}\right),\\
D'^E(\Theta S_{4}^-) &= \left(\begin{matrix} 
0 & 1 & 0 & 0 \\
-1 & 0 & 0 & 0 \\
0 & 0 & 0 & -1 \\
0 & 0 & 1 & 0
\end{matrix}\right).
\end{align}
Hence, we find that $D'^E$ is reducible,
\begin{equation}
D'^E = E \oplus E,
\end{equation}
with the two-dimensional real corep $E$. Next, we investigate basis functions where we again distinguish between basis functions that are odd with respect to $\Theta$, $ \psi_E^o=(\psi_{E1}^o,\psi_{E2}^o)$, and basis functions that are even, $\psi_E^e=(\psi_{E1}^e,\psi_{E2}^e)$. The basis functions of $D^2$ are calculated as
\begin{align}
	\gamma_E^o&=(\psi_E^o,\hat{P}_{\Theta S_4^+}\psi_E^o)=(\psi_{E1}^o,\psi_{E2}^o,-\psi_{E2}^o,\psi_{E1}^o),\\
	\gamma_E^e&=(\psi_E^e,\hat{P}_{\Theta S_4^+}\psi_E^e)=(\psi_{E1}^e,\psi_{E2}^e,\psi_{E2}^e,-\psi_{E1}^e).
\end{align}
For the basis functions of $D'^2$, we find 
\begin{align}
	\gamma'^o_E&= \sqrt{2}\, (\psi_{E1}^o, \psi_{E2}^o, 0, 0),\\
	\gamma'^e_E&= -\sqrt{2}\, (0, 0, \psi_{E1}^e, \psi_{E2}^e).
\end{align}
Hence, both $\psi_E^o$ and $\psi_E^e$ are basis functions of the real corep $E$. In Table \ref{table:char_tab_S4_C2}, the character table of $S_4(C_2)$ including the basis functions up to second order is given.

\section{Examples for band touchings}
\label{app.examples}

In this Appendix, we illustrate our method by analyzing the band touchings for three point groups: the monochromatic group $T$, the corresponding gray group, and the dichromatic group $D_4(C_4)$.

\subsection{Band touchings of the monochromatic point groups $T$ and the corresponding gray group}\label{app.examples_T}

\begin{figure*}
	\raisebox{34ex}[0ex][0ex]{(a)}\hspace{0.5em}\includegraphics[width=0.5\columnwidth]{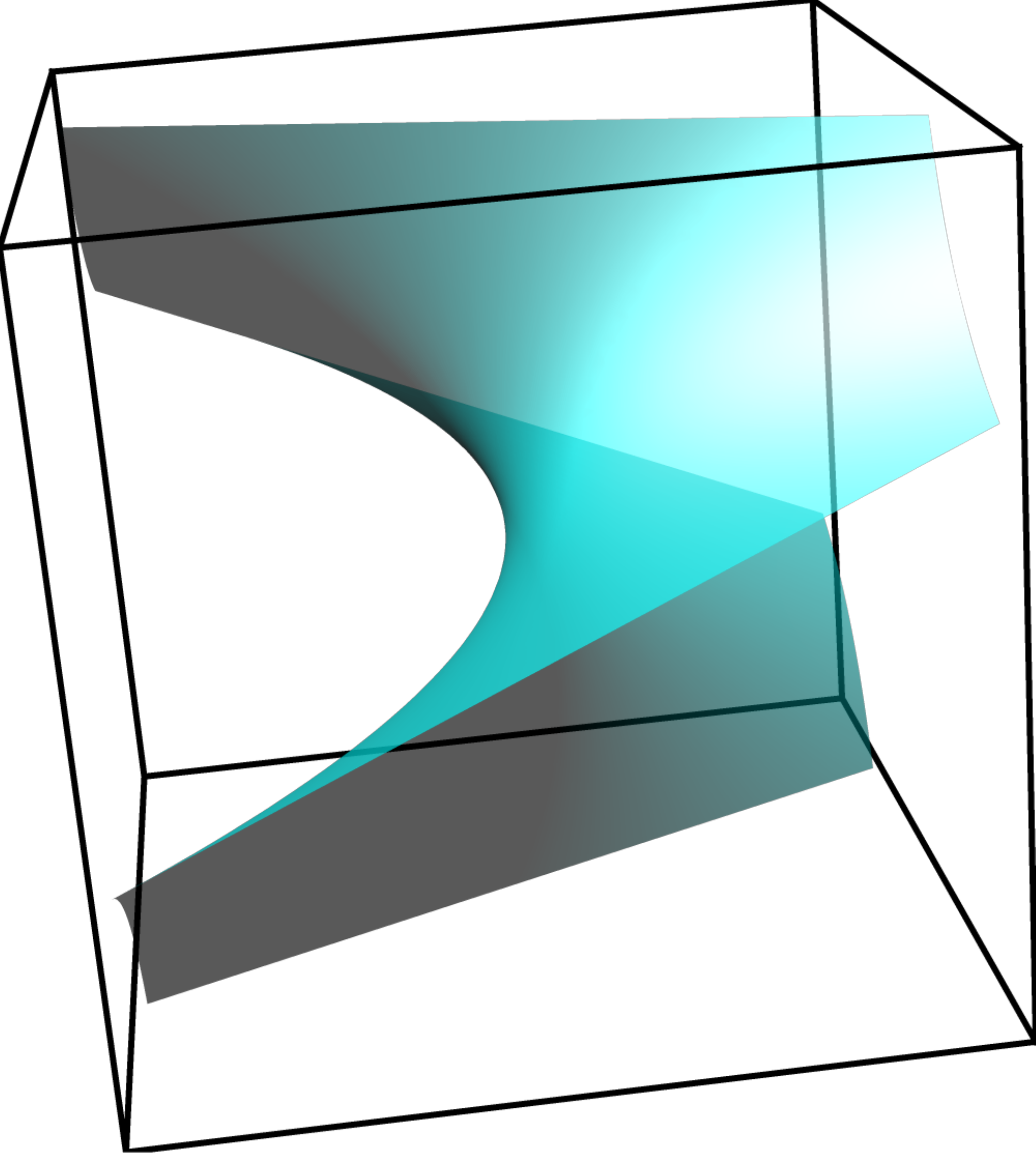}%
	\hspace{1.5em}\raisebox{34ex}[0ex][0ex]{(b)}\hspace{0.5em}\includegraphics[width=0.5\columnwidth]{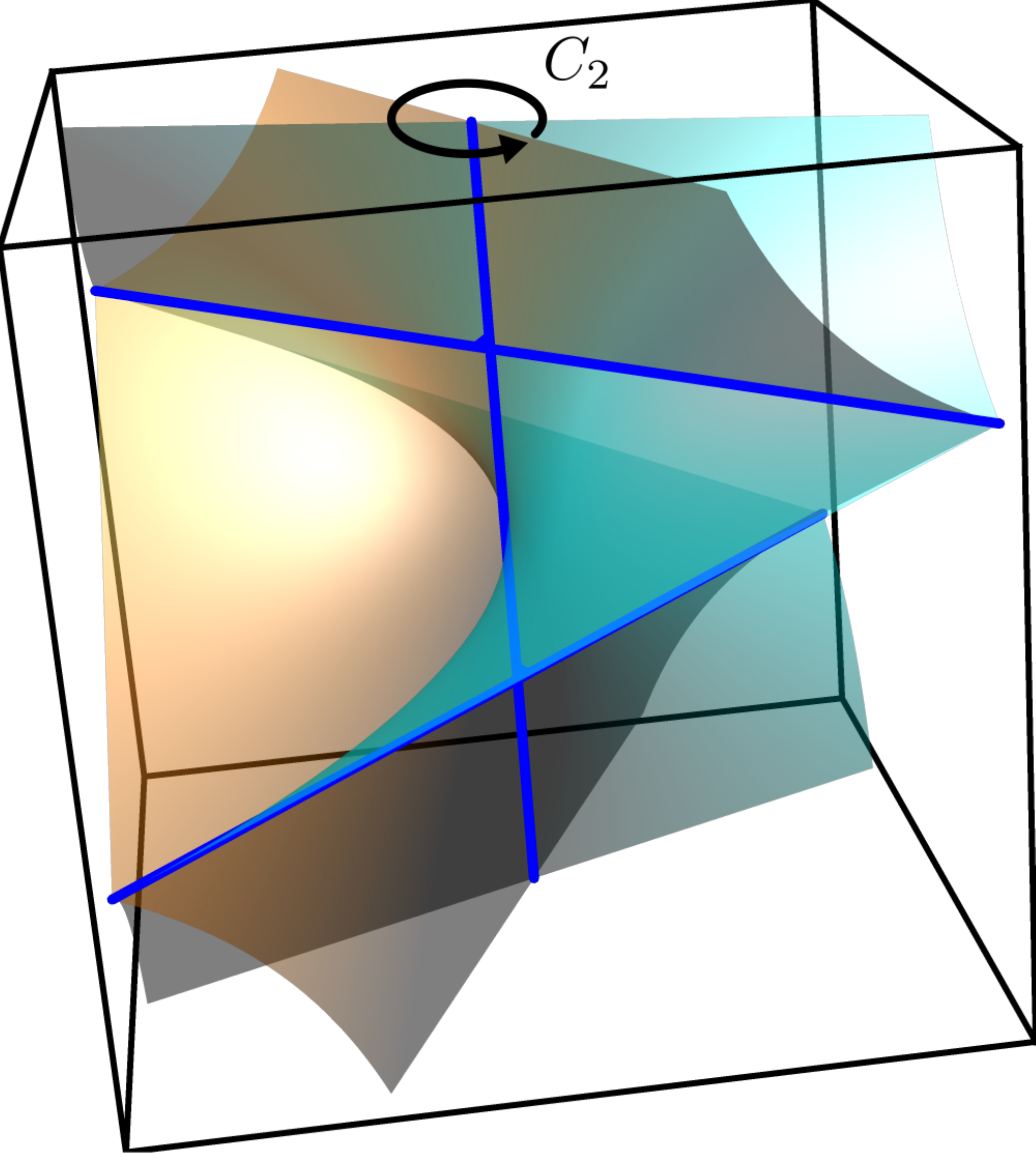}
	\hspace{1.5em}\raisebox{34ex}[0ex][0ex]{(c)}\hspace{0.5em}\includegraphics[width=0.5\columnwidth]{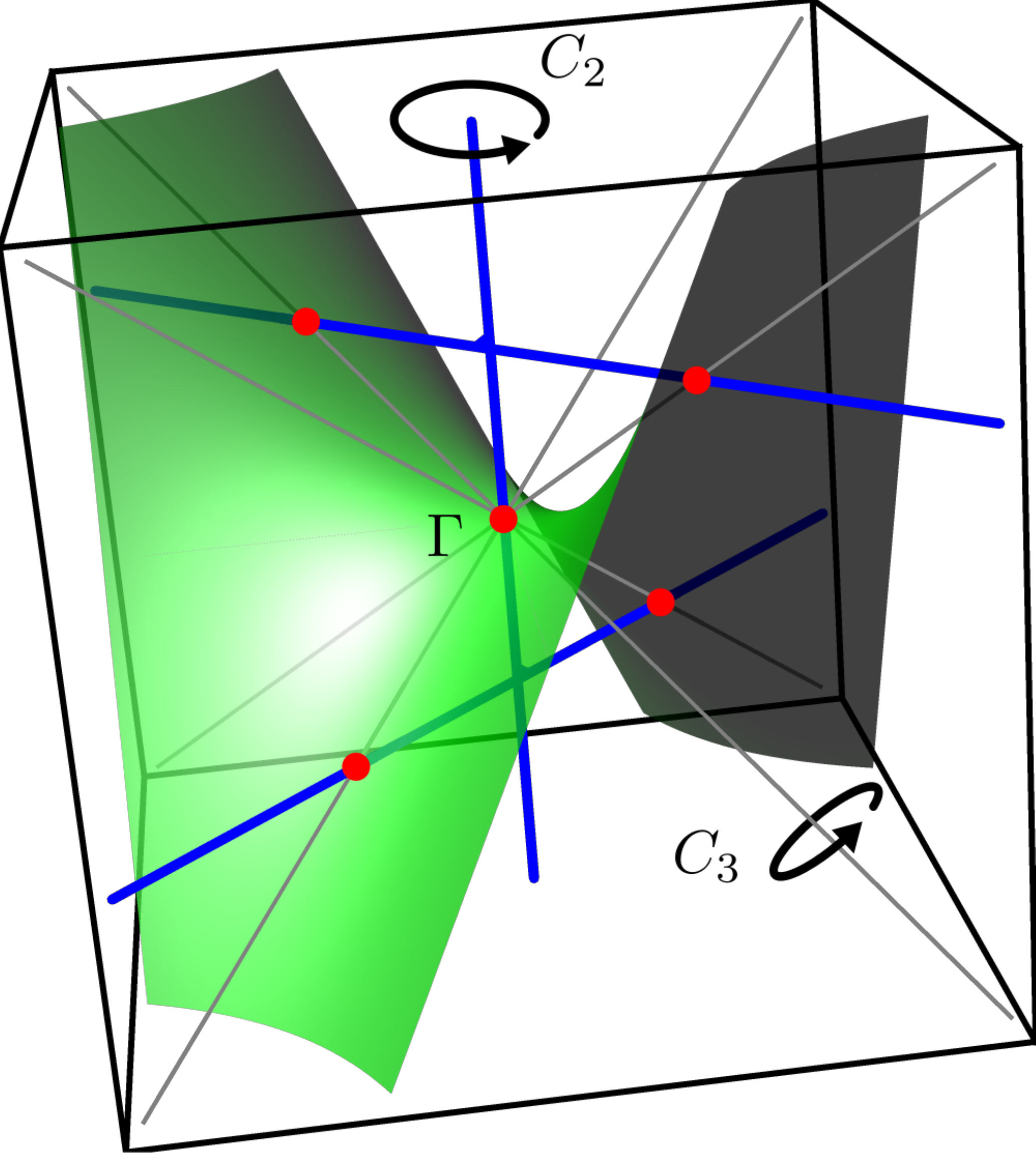}
	\caption{Solutions of the system of equations \eqref{eq:T_f_x}--\eqref{eq:T_f_z} for the monochromatic point group $T$. (a) The zero surface of $f_x^T=0$ is a hyperbolic paraboloid. (b) The common solutions of $f_x^T=0$ (light blue surface), and  $f_y^T=0$ (orange surface) are straight lines shown in dark blue. (c) The solutions of $f_x^T=f_y^T=0$ (dark blue) and $f_z^T=0$ (green surface) are given by Weyl points (red dots). The threefold rotation axes are shown in gray.}\label{fig_3}
\end{figure*}

In the following, we calculate the band touchings of the monochromatic point group $T$ close to the $\Gamma$ point. For the point group $T$, the Pauli matrices are elements of the three-dimensional representation $T$ \cite{Altmann_Herzig_1994}. Hence, we have to determine the zeros of the system of equations
\begin{align}
f^T_x(\bm{k})&= a_T\, k_x + b_T\, k_y k_z=0,\label{eq:T_f_x}\\
f^T_y(\bm{k})&= a_T\, k_y + b_T\, k_z k_x=0,\label{eq:T_f_y}\\
f^T_z(\bm{k})&= a_T\, k_z + b_T\, k_x k_y=0,\label{eq:T_f_z}
\end{align}
where $a_T, b_T \in \mathbb{R}$. Each of the equations \eqref{eq:T_f_x}--\eqref{eq:T_f_z} describes a hyperbolic paraboloid as illustrated in Fig.~\ref{fig_3}(a) for Eq.~\eqref{eq:T_f_x}. 

From Eq.~\eqref{eq:T_f_x}, we obtain 
\begin{equation}\label{eq:T_k_x}
k_x = - \frac{b_T\,k_y k_z}{a_T}. 
\end{equation}
We use this result to calculate the intersections of $f_x(\bm{k})=f_y(\bm{k})=0$ by plugging Eq.~\eqref{eq:T_k_x} into Eq.~\eqref{eq:T_f_y}:
\begin{equation}\label{eq:T_k_y}
f_y(\bm{k})=k_y\, \frac{a_T^2-b_T^2\, k_z^2}{a_T}=0. 
\end{equation}
The solutions of Eqs.\ \eqref{eq:T_k_x} and \eqref{eq:T_k_y} are calculated as 
\begin{equation}\label{eq:T_f_x_eq_f_y_1}
k_y = 0,\quad  k_x = 0,
\end{equation}
and
\begin{equation}\label{eq:T_f_x_eq_f_y_2}
k_z = \pm \frac{a_T}{b_T},\quad k_x = \mp k_y.
\end{equation}
The solution in Eq.\ \eqref{eq:T_f_x_eq_f_y_1} describes the entire $k_z$ axis, while the solution in Eq.\ \eqref{eq:T_f_x_eq_f_y_2} corresponds to two straight lines. These two lines lie in planes parallel to the $k_xk_y$ plane and are displaced in the normal direction by $k_z = \pm a_T/b_T$. They have polar angles $\mp \pi/4$ within these planes. The three different solutions of Eq.~\eqref{eq:T_k_y} are illustrated in Fig.~\ref{fig_3}(b).

\begin{figure*}
	\raisebox{34ex}[0ex][0ex]{(a)}\hspace{0.5em}\includegraphics[width=0.5\columnwidth]{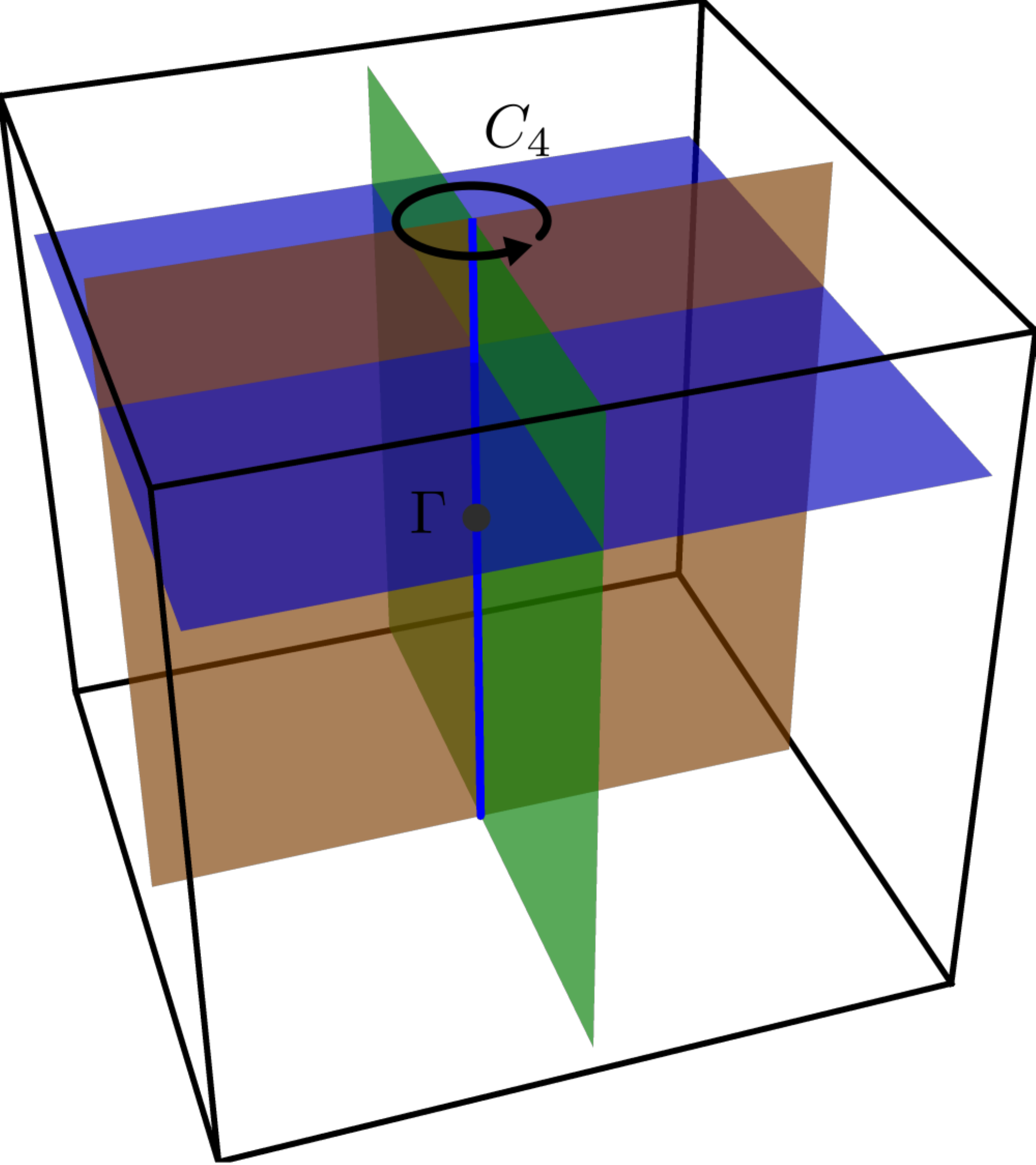}
	\hspace{1.5em}\raisebox{34ex}[0ex][0ex]{(b)}\hspace{0.5em}\includegraphics[width=0.5\columnwidth]{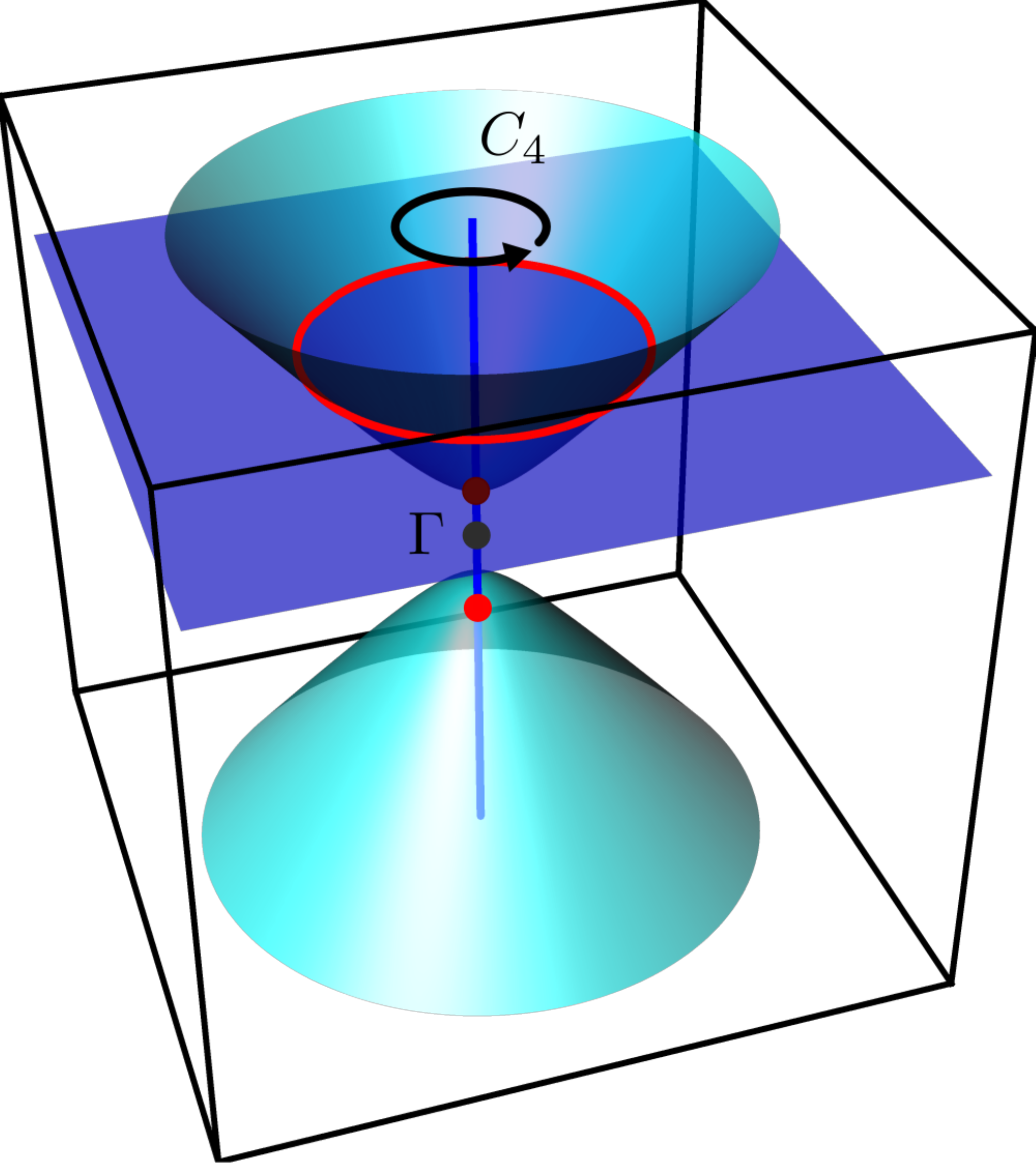}\\[2ex]
	\raisebox{34ex}[0ex][0ex]{(c)}\hspace{0.5em}\includegraphics[width=0.5\columnwidth]{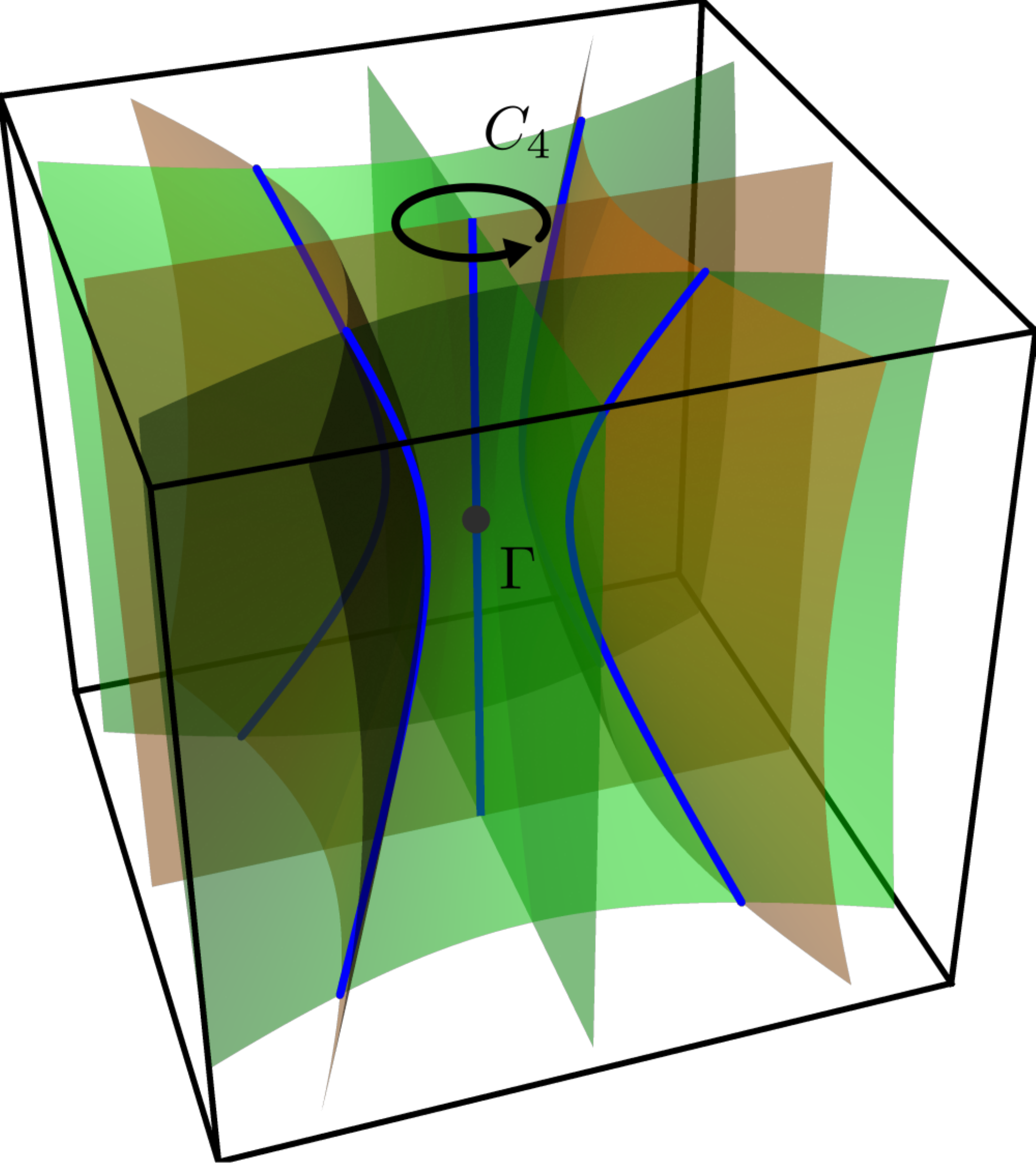}
	\hspace{1.5em}\raisebox{34ex}[0ex][0ex]{(d)}\hspace{0.5em}\includegraphics[width=0.5\columnwidth]{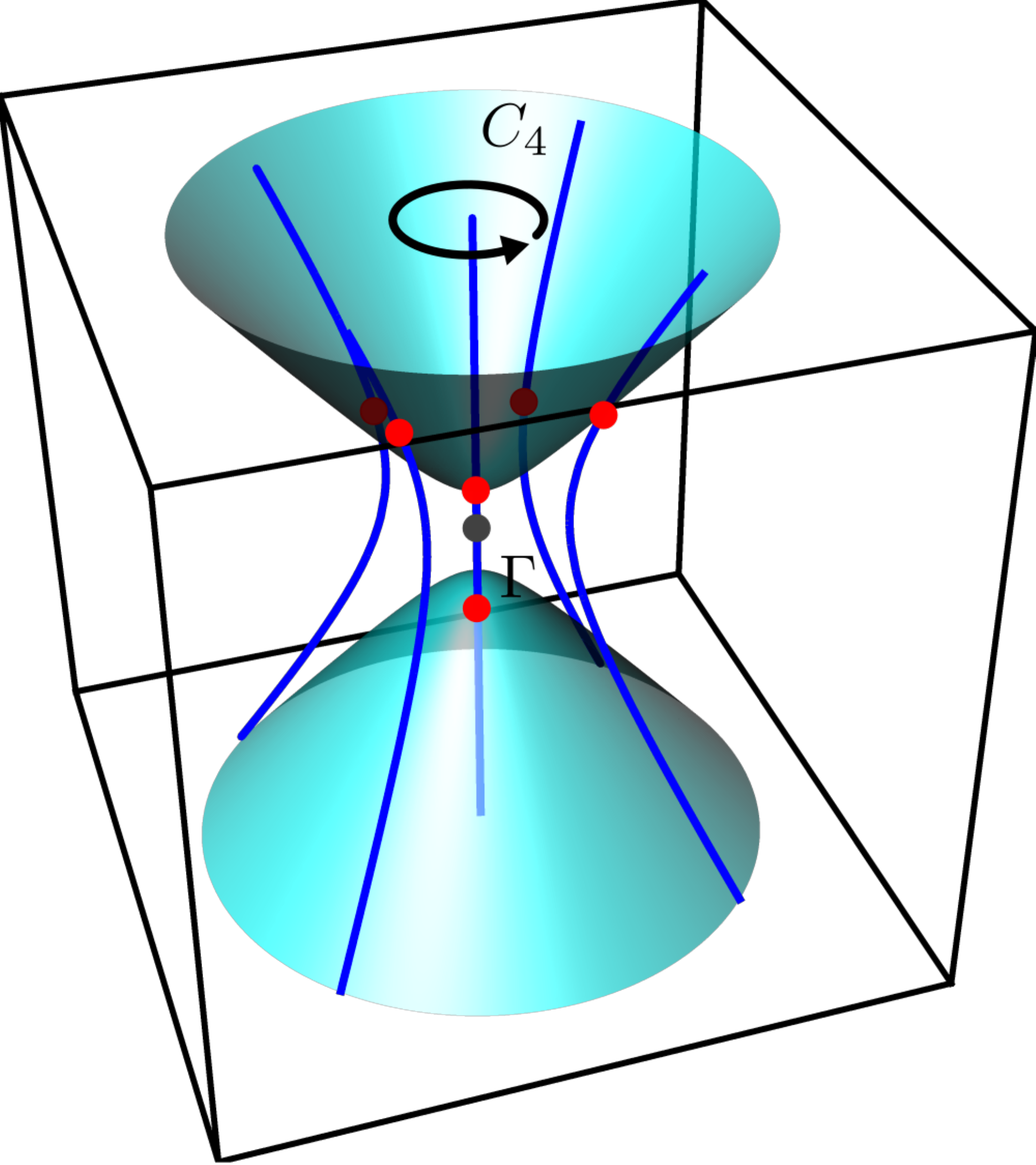}
	\caption{[(a), (b)] Solutions of the system of equations \eqref{eq:D4_C4_f_x}--\eqref{eq:D4_C4_f_z} for the dichromatic point group $D_4(C_4)$. (a) Common solution of $f_x^{\bm{M}}=0$ (green and blue surfaces) and $f_y^{\bm{M}}=0$ (orange and blue surface) in blue. (b) Common solution of $f_x^{\bm{M}}=f_y^{\bm{M}}=0$ (blue surface and line) and $f_z^{\bm{M}}=0$ (light-blue surfaces) in red. [(c), (d)] Solutions of the system of equations \eqref{eq:D4_C4_f_x_ho}, \eqref{eq:D4_C4_f_y_ho}, and \eqref{eq:D4_C4_f_z} where higher-order terms are taken into account. (c) Common solution of $\tilde{f}_x^{\bm{M}}=0$ (green surfaces) and $\tilde{f}^{\bm{M}}_{y} = 0$ (orange surfaces) in blue. Common solution of $\tilde{f}^{\bm{M}}_{x} = \tilde{f}^{\bm{M}}_y = 0$ (blue curves) and $f^{\bm{M}}_z =0$ (light-blue surfaces) in red.}\label{fig_4}
\end{figure*}

In the third step, we search for solutions of $f^T_x=f^T_y=f^T_z=0$. The solutions are illustrated in Fig.~\ref{fig_3}(c). By plugging the solution of $f^T_x=f^T_y=0$ given by Eq.~\eqref{eq:T_f_x_eq_f_y_1} into Eq.~\eqref{eq:T_f_z}, we find the single solution
\begin{equation}
\bm{k}^T_0 = (0,0,0), 
\end{equation}
which is the $\Gamma$ point. By inserting the solution in Eq.~\eqref{eq:T_f_x_eq_f_y_2} into Eq.~\eqref{eq:T_f_z}, we obtain the four independent solutions
\begin{align}
k^T_{1,2;z} = \frac{a_T}{b_T},\; k^T_{1,2;y} = \pm \frac{a_T}{b_T},\; k^T_{1,2;x} = - k^T_{1,2;y} = \mp \frac{a_T}{b_T}
\end{align}    
and
\begin{align}
k^T_{3,4;z} = -\frac{a_T}{b_T},\; k^T_{3,4;y} = \pm \frac{a_T}{b_T},\; k^T_{3,4;x} = + k^T_{3,4;y} =  \pm \frac{a_T}{b_T}. 
\end{align}
In total, there are five Weyl points as illustrated in Fig.~\ref{fig_3}(c). One Weyl point is the $\Gamma$ point, while the other four are located on the four threefold rotation axes.

Next, we investigate which Weyl points remain if we take into account time-reversal symmetry. For the gray point group $T\otimes\{e,\Theta\}$, the Pauli matrices and thus the coefficient functions $f^T_i$ belong to the time-reversal-odd real corep $T^-$. Hence, only the odd-in-momentum basis functions are relevant for the coefficient functions and we have to solve
\begin{align}
\tilde{f}^T_x(\bm{k})&= a_T\, k_x=0,\label{eq:T_g_f_x}\\
\tilde{f}^T_y(\bm{k})&= a_T\, k_y=0,\label{eq:T_g_f_y}\\
\tilde{f}^T_z(\bm{k})&= a_T\, k_z=0,\label{eq:T_g_f_z}
\end{align}
where we have included all momentum basis functions up to second order. The only solution of this system of equations is $\bm{k}=(0,0,0)$. Consequently, out of the five Weyl points of the monochromatic point group $T$, only the Weyl point at the time-reversal-invariant $\Gamma$ point survives. 

\subsection{Band touchings of the dichromatic point group $D_4(C_4)$}
\label{app.example.D4C4}

Here, we investigate the band touchings of the dichromatic point group $\bm{M}=D_4(C_4)$, the character table of which is given in Table \ref{table:char_tab_D4_C4} above. We have to calculate the solutions of the system of equations 
\begin{align}
f^{\bm{M}}_x(\bm{k})&=g_{\bm{M}}\, k_x + h_{\bm{M}}\, k_x k_z=0, \label{eq:D4_C4_f_x}\\
f^{\bm{M}}_y(\bm{k})&=g_{\bm{M}}\, k_y + h_{\bm{M}}\, k_y k_z=0, \label{eq:D4_C4_f_y}\\
f^{\bm{M}}_z(\bm{k})&=a_{\bm{M}}+ b_{\bm{M}}\, k_z + c_{\bm{M}}\, (k_x^2+k_y^2) + d_{\bm{M}}\, k_z^2=0, \label{eq:D4_C4_f_z}
\end{align}   
where $a_{\bm{M}}, b_{\bm{M}}, c_{\bm{M}}, d_{\bm{M}}, g_{\bm{M}}, h_{\bm{M}} \in \mathbb{R}$. The intersections of the surfaces $f^{\bm{M}}_x=f^{\bm{M}}_y=0$ described by Eqs.~\eqref{eq:D4_C4_f_x} and \eqref{eq:D4_C4_f_y} are given by the $k_z$ axis and a plane parallel to the $k_x k_y$ plane, as illustrated in Fig.~\ref{fig_4}(a). Equation \eqref{eq:D4_C4_f_z} describes a quadric the center of which is on the $k_z$ axis. Furthermore, the principal axes of this quadric are the coordinate axes and the $k_x$ and $k_y$ principal axes are degenerate. Thus, the solutions of $f^{\bm{M}}_x=f^{\bm{M}}_y=f^{\bm{M}}_z=0$ are either zero or two Weyl points on the $k_z$ axis, which is the fourfold rotation axis, and there can also be a circular nodal line the center of which is on the $k_z$ axis. The circular nodal line is located in a plane that is parallel to the $k_x k_y$ plane. In Fig.~\ref{fig_4}(b), one possible solution of the system of equations \eqref{eq:D4_C4_f_x}--\eqref{eq:D4_C4_f_z} is depicted for parameters for which $f^{\bm{M}}_z=0$ describes a hyperboloid of two sheets.

Next, we investigate the stability of the circular nodal line by taking into account third-order terms for $f^{\bm{M}}_x$ and $f^{\bm{M}}_y$. The corresponding equations then read as
\begin{align}
\tilde{f}^{\bm{M}}_x(\bm{k}) &= g_{\bm{M}}\, k_x + h_{\bm{M}}\, k_x k_z + m_{\bm{M}} \, k_x k_z^2 \nonumber \\
&\quad{}+ n_{\bm{M}}\, k_x (k_x^2-3k_y^2) = 0, \label{eq:D4_C4_f_x_ho}\\
\tilde{f}^{\bm{M}}_y(\bm{k}) &= g_{\bm{M}}\, k_y + h_{\bm{M}}\, k_y k_z + m_{\bm{M}} \, k_y k_z^2 \nonumber \\
&\quad{}+ n_{\bm{M}}\, k_y (k_y^2-3 k_x^2) = 0, \label{eq:D4_C4_f_y_ho}
\end{align}
where $m_{\bm{M}}, n_{\bm{M}} \in \mathbb{R}$. From Eqs.~\eqref{eq:D4_C4_f_x_ho} and \eqref{eq:D4_C4_f_y_ho}, we find that $\tilde{f}^{\bm{M}}_x=\tilde{f}^{\bm{M}}_y=0$ is still solved by all points on the $k_z$ axis. However, there is no longer a solution in the form of a plane parallel to the $k_x k_y$ plane. Instead, there are four curves in addition to the $k_z$ axis, as depicted in Fig.~\ref{fig_4}(c). Thus, the circular line node in Fig.\ \ref{fig_4}(b) dissolves into Weyl points when we take into account higher order terms, as depicted in Fig.~\ref{fig_4}(d). The line node found to second order is thus fragile.

\bibliography{Knoll_literature}

\end{document}